\documentclass[letterpaper,twocolumn,aps,prl,floatfix,
  citeautoscript,nobibnotes]{revtex4-1} 
\usepackage{epsfig,epstopdf}
\usepackage{graphicx}
\usepackage{amsfonts,amsmath,amssymb}
\usepackage{latexsym}
\usepackage{bm}
\usepackage{color}
\usepackage[colorlinks=true,allcolors=blue,breaklinks=true]{hyperref}

\newcommand*\chem[1]{\ensuremath{\mathrm{#1}}}

\begin{document}

\title{Monte Carlo study on the detection of classical order by disorder in real antiferromagnetic Ising pyrochlores}

\author{Pamela C. Guruciaga}
\email[Corresponding author: ]{pamela.guruciaga@cab.cnea.gov.ar}
\affiliation{Centro At\'omico Bariloche, Comisi\'on Nacional de Energ\'{\i}a At\'omica (CNEA), Consejo Nacional de Investigaciones Cient\'{\i}ficas y T\'ecnicas (CONICET), Av. Exequiel Bustillo 9500, R8402AGP San Carlos de Bariloche, R\'{\i}o Negro, Argentina}

\author{Rodolfo A. Borzi}
\affiliation{Instituto de F\'{\i}sica de L\'{\i}quidos y Sistemas Biol\'ogicos (IFLYSIB), UNLP-CONICET, B1900BTE La Plata, Buenos Aires, Argentina}
\affiliation{Departamento de F\'{\i}sica, Facultad de Ciencias Exactas, Universidad Nacional de La Plata, c.c. 16, suc. 4, B1900AJL La Plata, Argentina}


\begin{abstract}
    We use Monte Carlo simulations to evaluate the feasibility of detecting thermal order by disorder in real antiferromagnetic Ising pyrochlores, frustrated by a magnetic field applied in the $[110]$ direction.  
    Building on an ideal system with only nearest-neighbour exchange interactions and a perfectly oriented field, we consider the effects of dipolar interactions and field misalignment. Our approach is special in that it relies more in the possibility to switch on and off the entropic drive towards order than in the absence of (or immunity to) a particular perturbation. It can then be applied, in principle, to other uncontrolled interactions expected to be naturally present in real {magnetic} materials.
    We establish the conditions under which entropic effects can be discerned from an interaction drive towards order, show how to use neutron scattering as a means to unveil this mechanism, and discuss possible materials where to test these ideas.
\end{abstract}

\maketitle

\section{Introduction}

In magnetism, the geometry of the spin lattice can be just an unimportant detail (as when considering critical fluctuations in a ferromagnet~\cite{Cardy1996scaling}) or a crucial one, deciding whether magnetic order actually takes place or not. The latter case applies to geometrically-frustrated magnetic systems, characterised by the absence of spin order at low temperatures even without any quenched structural disorder~\cite{Ramirez1994review,Moessner2006review}. Prominent examples among these disordered systems are the magnetic equivalents of a liquid ({\textit{spin} liquids,} correlated but disordered)~\cite{Balents2010spin,Lacroix2011introduction} and of ice {(\textit{spin} ices)}~\cite{Bramwell01,Diep2013frustrated}. They are home to very exotic quasiparticles~\cite{Balents2010spin,Fennell2014magnetoelastic} and to the magnetic analogue of electric charges or ``monopoles'', respectively. The monopoles interact through a Coulomb law and respond to a magnetic field like charges to an electric one~\cite{Castelnovo}. In { further} analogy to systems of electric charges, different states of \emph{monopole matter} have been suggested or detected: monopole fluids~\cite{Slobinsky2018charge,Borzi13,Guruciaga14}, ionic-like crystals with \emph{staggered} magnetic-charge density~\cite{Castelnovo,Borzi13,Brooks14,Guruciaga14,jaubert2015holes}, or much more complicated structures~\cite{Sazonov2012double,Jaubert2015crystallog,Borzi2016intermediate,Udagawa2016,Rau2016spinslush}. Much of {these} physics can be extended into two-dimensional artificial structures~\cite{Xie2015magnetic,Farhaneaav6380,Olson_Reichhardt_2012}, opening also to new possibilities.

Since strong spin correlations are operative, in principle even very small perturbations can break the massive degeneracy of some of the fluid-like (or even crystalline~\cite{Borzi13,Brooks14}) ground states, forcing the appearance of long-range order at low-enough temperature. Adding to this, the fact that the ground state multiplicity in these frustrated systems is accidental
opens up the possibility to other --more subtle-- mechanisms of symmetry breaking. Seemingly against thermodynamics, even thermal disorder can be the driving force helping to develop an order parameter. Mirroring this apparent contradiction, the phenomenon was baptised as classical or \emph{thermal} ``order by disorder''~\cite{Chalker,moessner1998low,Shender1996review,bramwell1994order} (OBD). It was first shown by Villain and collaborators in the so called generalised \textit{domino model}~\cite{Villain}, and it occurs when there is a huge disproportion in the density of low-energy excitations associated with a particular ground state.

Theoretical models where disorder (including even structural defects, see for example Ref.~\onlinecite{Maryasin2014structural}) helps to develop a magnetic order parameter are quite interesting by themselves and have been studied for many years~\cite{Villain,Henley1989ordering}. However, a crucial question to ask when dealing with real systems is how to determine that the main cause of order is related to the entropy term ($-TS$) of the free energy $F = U_0 - TS$, and not to an uncontrolled energy perturbation affecting the internal energy $U_0$ (see discussion on Ref.~\onlinecite{Savary12}). The strongest case in favour of actual OBD at the moment seems to be that of the XY pyrochlore \chem{Er_2Ti_2O_7}, where the classical ground state degeneracy is protected by symmetry~\cite{Savary12}. The actual mechanism is still under discussion~\cite{Rau2016order}, and quantum~\cite{Savary12,Zhitomirsky12,Ross14,Champion03}, classical~\cite{Mcclarty2014order} or composed~\cite{Oitmaa13} fluctuations could be the selection mechanism behind the observed order.

In this paper we explore in detail this crucial point of the detection of OBD in real materials. Following Ref.~\onlinecite{Guruciaga16}, we will see that antiferromagnetic Ising pyrochlores~\cite{Sleight1974semiconductor,Lhotel2015fluctuations,Xu2015,Anand15,Bertin2015nd} are a promising experimental scenario to detect OBD. While it is never possible to switch off every term affecting the unperturbed internal energy $U_0$ in a real crystal, the physics on these materials naturally suggest the reverse approach: the thermally-driven tendency towards magnetic charge order in the $-TS$ term can be detuned by means of an easily-controllable parameter. In this way, we manipulate the magnetic field to change the type of excitations occurring at fixed temperature, gauging the importance of the OBD mechanism by its effect on the staggered charge density. Using Monte Carlo simulations, we apply this method to the particular case of dipolar interactions. These, aside from being expected in real materials, allow understanding the perturbation as an attraction among Coulomb charges and thus favouring charge order. In principle, the same ideas could be applied to any other perturbation terms affecting $U_0$ (e.g., second or further neighbours exchange interactions), {or to other systems presenting field-tuned OBD (e.g., artificial spin ices~\cite{Guruciaga16}, where dipolar interactions are unavoidable)}.
We also study the effect of a misaligned magnetic field $\mathbf{B}$. This case is special since the control parameter used to put in evidence the OBD mechanism is (in addition to this) a force driving the order{; in other words, $\textbf{B}$ changes $S(T,\mathbf{B})$ in a nontrivial way but also perturbs $U_0$}.
With the intention to bridge the gap between theory and experiment, we suggest materials where OBD could be detected using these ideas and how to measure the effect, concentrating in the particular case of neutron scattering techniques.


\subsection{System and model}

\begin{figure*}[t]
    \centering
    \includegraphics[width=\linewidth]{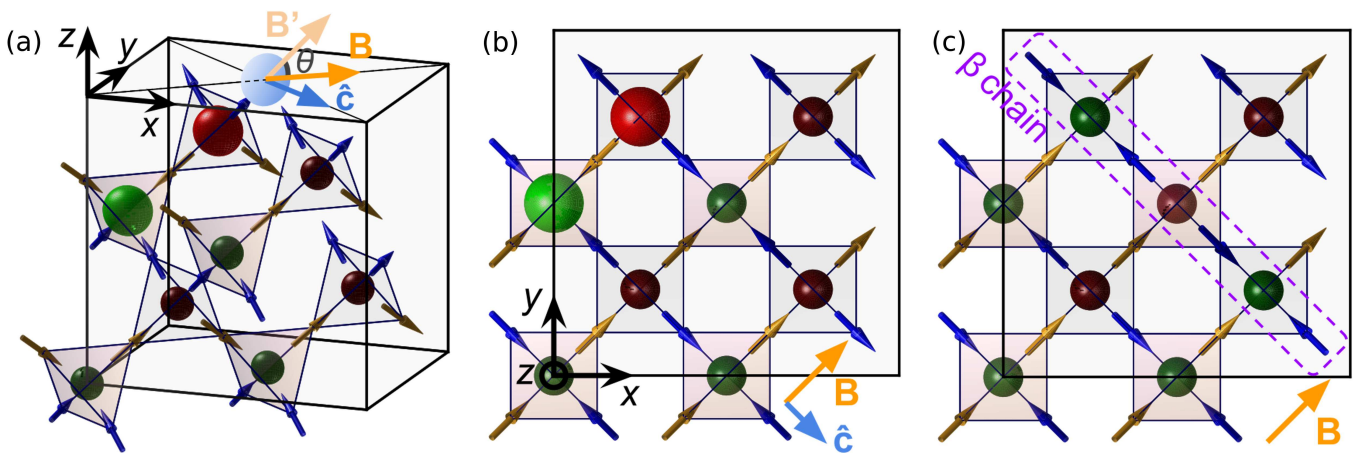}
    \caption{(a)~Conventional unit cell of the pyrochlore lattice ($L=1$), consisting of up (coloured) and down (uncoloured) tetrahedra. Spins (arrows) sit in their vertices and can be mapped into magnetic charges or \emph{monopoles} (spheres, the colour and size of which represent their sign and value). {The magnetic field $\mathbf{B}\parallel [110]$ can be inadvertently misdirected by a small angle $\theta$ around $\hat{\mathbf{c}}\propto [1\overline{1}0]$ to give a tilted field $\mathbf{B}'$ with a nonzero component along $[001]$. (b)~The same configuration, which presents staggered charge order, viewed from top. While $\beta$ spins (blue) are orthogonal to the perfectly-oriented magnetic field $\mathbf{B}$, $\alpha$ spins (yellow) couple with it.} (c)~Disordered zero-temperature ground state for $\mu B =h>h_{\textit{inf}}$ in the ideal $D=0,\, \theta=0$ case. Incoherent single-charge-ordered $\beta$ chains (running perpendicular to $\mathbf{B}$) populate the system and produce a globally-disordered state with $\rho^s=1$ and $\rho_S^s=0$.}
    \label{piroc}
\end{figure*}

The pyrochlore lattice consists of corner-sharing tetrahedra, the centres of which form a diamond lattice (Fig.~\ref{piroc}(a)). Classical Ising magnetic moments ${\boldsymbol \mu}_i =\mu {\mathbf S}_i  =\mu S_i \hat{\mathbf s}_i $ sit on the vertices of the tetrahedra, with the quantisation directions $\hat {\mathbf s}_i$ along the local $\langle 111 \rangle$ directions. The pseudospins $S_i = \pm 1$ indicate if the magnetic moments point outwards ($+1$) or inwards ($-1$) of  ``up'' tetrahedra (coloured ones in Fig.~\ref{piroc}(a)). The Zeeman coupling with a magnetic field $\mathbf{B}$, as well as the magnetic interactions of exchange and dipolar origin, of strengths $J$ and $D$, respectively, are accounted for by the Hamiltonian
\begin{align}\label{eq:dsim}
    {\mathcal H} =& -J\sum_{\langle ij\rangle}{\mathbf S}_i \cdot{\mathbf S}_j - \mu \sum_i {\mathbf B} \cdot {\mathbf S}_i   \nonumber\\
        &+ D\, r_{\textit{nn}}^3 \sum_{i>j} \left[ \frac{{\mathbf S}_i \cdot{\mathbf S}_j}{|{\bf{r}}_{ij}|^3}- \frac{3({\mathbf S}_i \cdot {\bf{r}}_{ij}) ({\mathbf S}_j \cdot {\bf{r}}_{ij}) }{|{\bf{r}}_{ij}|^5} \right] \, ,
\end{align}
where $\langle ij\rangle$ means the sum is carried over nearest neighbours, $r_{\textit{nn}}$ is { the pyrochlore nearest-neighbour distance} and $D=\mu_0\mu^2/(4\pi r_{\textit{nn}}^3)$. 
It can be easily seen in Fig.~\ref{piroc} that the {$N$} spins may be separated into two groups according to how they couple to a field $\mathbf{B}\parallel[110]$: while $\beta$ spins (blue) are orthogonal to it, $\alpha$ spins (yellow) have a nonzero projection ${\hat {\mathbf s}}_i \cdot {\mathbf B} = \alpha_i \sqrt{2/3}\, B$ with $\alpha_i = \pm 1$~\cite{Hiroi03}. Using these definitions, and referring all energy contributions to the nearest-neighbour interactions term, the Hamiltonian can be rewritten as
\begin{equation}
    \frac{{\mathcal H}}{|J_{\textit{eff}}| } = -\sum_{\langle ij\rangle} S_i S_j - \frac{\sqrt{2}}{\sqrt{3}} \frac{h}{|J_{\textit{eff}}|} \sum_{i\in \alpha} \alpha_i S_i  + {\frac{D}{|J_{\textit{eff}}| }} \mathcal{H}_{\textit{dip}}^{r>r_{\textit{nn}}}\; ,
    \label{eq:dsim2}
\end{equation}
with $J_{\textit{eff}}=J/3+5D/3$ (assumed $<0$ throughout this text), and $h=\mu B$, { with the second sum running over $\alpha$ spins only}. $\mathcal{H}_{\textit{dip}}^{r>r_{\textit{nn}}}$ {is unitless, and} encompasses all dipolar contributions beyond nearest neighbours. { The ferromagnetic version of Eq.~\eqref{eq:dsim2} corresponds to the well-known dipolar spin-ice model~\cite{Melko}. In our work we will focus on { a case in which the antiferromagnetic nearest-neighbour term ($J_{\textit{eff}}<0$) modified by the Zeeman energy dominates. As shown analytically and numerically in Ref.~\citenum{Guruciaga16}, Eq.~\eqref{eq:dsim2} with $D/|J_{\textit{eff}}|=0$ conducts to physics similar to Villain's domino model~\cite{Villain}, but its field-tuned OBD may allow for an experimental contrast of the phenomenon. Different from this recent work, the nearest neighbor Hamiltonian is now} perturbed by dipolar interactions $\mathcal{H}_{\textit{dip}}^{r>r_{\textit{nn}}}$}, with a relative intensity controlled by the (assumed small) parameter $D/|J_{\textit{eff}}|$.

Within the monopole picture~\cite{Castelnovo}, the diverse spin configurations can be mapped to arrays of magnetic charges that live in the dual diamond lattice, {with a lattice constant $r_d=\sqrt{3/2}\, r_{\textit{nn}}$}. While tetrahedra with two spins pointing in and two out are considered neutral, those with three spins in and one out, or viceversa, have a positive or negative single charge $Q_s=2\mu/r_d$ sitting in their centres (small spheres in Figs.~\ref{piroc}(a)-(c)). Moreover, positive or negative double charges (big spheres in Figs.~\ref{piroc}(a) and~\ref{piroc}(b)), with $Q_d=2Q_s$, can be found in tetrahedra with all spins pointing in or all spins pointing out, {respectively}. It will be useful to define the density of single monopoles $\rho^s$, which has a value of $1$ when there is exactly one single monopole per tetrahedron; analogously, one can define the density of double monopoles $\rho^d$ and neutral sites $\rho^n$.~\footnote{Note that, differently from the staggered charge density defined below, these are \emph{number} densities, irrespective of the monopole charges (which always average to zero).} Having transformed a configuration of spins into one of monopoles, dipolar interactions can in turn be approximately described as Coulombian forces between monopoles, as was shown by Castelnovo and collaborators~\cite{Castelnovo}. Since these forces promote the development of ordered phases of charges of alternating sign with the structure of zinc blende~\cite{Borzi13,Brooks14,Guruciaga14}, we define here two order parameters: the single-monopole (double-monopole) staggered \emph{charge} density $\rho_S^s$ ($\rho_S^d$) as the {thermal} average of the modulus of the {\it total} magnetic charge due to single (double) charges in up tetrahedra, normalised so that full order corresponds to a value of $1$. 
{Mathematically, the single-monopole staggered density can be written}
\begin{equation}
    \rho_S^s=\bigg\langle\frac{1}{N_{\textit{up}}}\left| \sum_{\nu} a_{\nu}^s \right|\bigg\rangle \, , 
\end{equation}
where the Greek index $\nu$ denotes a sum carried over the \textit{diamond}-lattice sites corresponding to the centres of up tetrahedra (the total number of which is $N_{\textit{up}}=N/4$), and $a_{\nu}^s$ is equal to $1$ ($-1$) if there is a positive (negative) single monopole in site $\nu$, and equal to $0$ otherwise. {Analogously, the double monopole staggered density is defined as
\begin{equation}
    \rho_S^d=\bigg\langle\frac{1}{N_{\textit{up}}}\left| \sum_{\nu} a_{\nu}^d \right|\bigg\rangle \, , 
\end{equation}
with $a_{\nu}^d$ equal to $1$ ($-1$) if a positive (negative) double monopole sits in site $\nu$ of the up tetrahedra sublattice, and equal to $0$ otherwise.}
According to these definitions, having for example {$\rho_S^s < 1$} can imply both a lattice full of partially disordered single charges, as well as one with a low density of single charges but with a perfect staggered order. { The measurement of the density of single and double charges erases any ambiguity.}


\subsection{Simulation details}

{We performed Monte Carlo simulations with the Metropolis algorithm and single-spin-flip dynamics. In order to implement Eq.~\eqref{eq:dsim2} in the algorithm, we used Ewald summations to take into account long-range interactions. Different materials were explored by changing the adimensional parameters in the Hamiltonian. In practice, we varied the value of $D$ (for example, increasing it to consider a material with a big $\mu$) modifying $J$ accordingly so as to keep $J_{\textit{eff}}$ constant. $B$ was always varied freely, from which we obtained the product $h=\mu B$.}

The conventional unit cell of the pyrochlore lattice (Fig.~\ref{piroc}(a)) consists of $16$ spins, and we simulated cubic systems of $L^3$ cells with periodic boundary conditions. As an example, a system with $L=6$ and dipolar interactions took $5\times 10^4$ Monte Carlo steps to reach the equilibrium, and then $2.5\times 10^4$ steps were needed to calculate the averages at each value of temperature and applied magnetic field. In turn, the results were averaged over $30$ independent runs. 

We used spin configurations obtained from Monte Carlo runs to simulate the spin structure factor given by
\begin{equation} \label{spinSk}
    I_{{\textit{SS}}}(\boldsymbol{k})=\frac{1}{N}\sum_{i=1}^N \sum_{j=1}^N \langle S_i S_j\rangle \left(\hat {\mathbf s}_i^{\perp}\cdot\hat {\mathbf s}_j^{\perp} \right) e^{i\boldsymbol{k}\cdot \boldsymbol{r}_{ij}} \, ,
\end{equation}
where $N$ is the number of sites in the pyrochlore lattice, $\langle S_i S_j\rangle$ is the thermal average of the correlation between pseudospins at sites $i$ and $j$, and $\hat {\mathbf s}_i^{\perp}$ is the component of the quantisation direction at site $i$ perpendicular to the scattering wave vector $\boldsymbol{k}$, calculated as
\begin{equation}
    \hat{\mathbf s}_i^{\perp}=\hat {\mathbf s}_i -\left(\hat {\mathbf s}_i\cdot\frac{\boldsymbol{k}}{|\boldsymbol{k}|}\right) \frac{\boldsymbol{k}}{|\boldsymbol{k}|} \, .
\end{equation}
We also simulated the charge-charge correlation function by following the expression
\begin{equation}\label{chargeSk}
    I_{{\textit{QQ}}}(\boldsymbol{k})=\frac{2}{N}\sum_{\eta=1}^{N/2}\,\sum_{\nu=1}^{N/2}\langle Q_{\eta} Q_{\nu}\rangle\,e^{i\boldsymbol{k}\cdot \boldsymbol{r}_{\eta\nu}}\, ,
\end{equation}
where $N/2$ is the number of tetrahedra, $Q_{\eta}$ represents the topological charge at site $\eta$ of the diamond lattice, and ${\boldsymbol{r}}_{\eta\nu}$ is the distance between monopoles.
Both the spin structure factor and the charge-charge correlation function were obtained by averaging over sets composed of $150-200$ configurations for $L=8$.


\section{Effect of dipolar interactions}
\subsection{Ground state and low-energy excitations}\label{GSandexcitations}

\begin{figure}[t]
    \centering
    \includegraphics[width=1\linewidth]{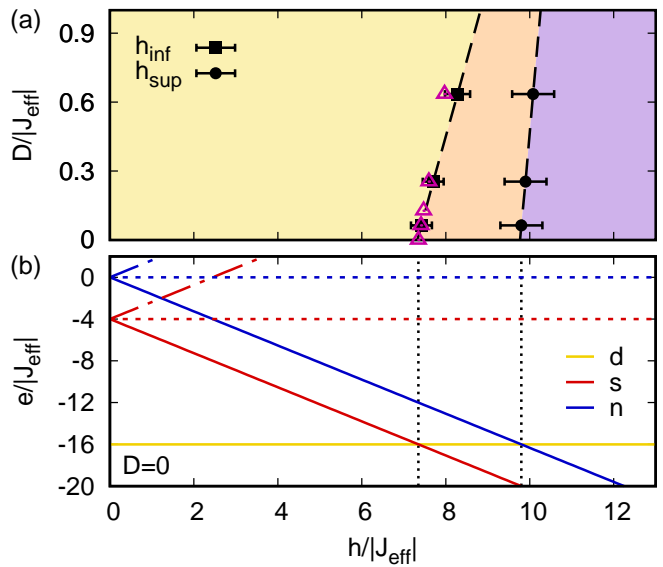}
    \caption{(a)~{{Diagram showing the two phases of} the antiferromagnetic Ising pyrochlore with dipolar interactions of strength $D$ under a magnetic field $h$ in the $[110]$ direction. The yellow sector marks the double-monopole~(\textit{d}) ground state. {The sectors painted $\text{orange}=\text{red}+\text{yellow}$ (where OBD can take place) and $\text{purple}=\text{red}+\text{blue}$ correspond to the same thermodynamic phase: a single-monopole~(\textit{s}) ground state, with double charges or neutral sites~(\textit{n}) as the {lowest-energy} excitations, respectively. The two ground states are separated by $h_{\textit{inf}}$ (full line), while the change in the type of favoured excitations at low temperature is indicated by $h_{\textit{sup}}$ (dashed line).} Both fields were obtained by simulating spin systems (full squares and circles) {in the $T\to 0$ limit (see end of Sec.~\ref{GSandexcitations} for details)}; $h_{\textit{inf}}$ is also compared to the values calculated within the monopole picture (open triangles). (b)~Nearest-neighbour ($D=0$) energy of the different spin configurations of a single tetrahedron as a function of the field. The colours used for the different types of charge are in correlation with the panel above. During this work we will be interested only in the lowest-energy configurations of each kind (full lines). In this panel, the region where OBD can occur is marked by vertical dashed lines.}}
    \label{E-D_vs_B}
\end{figure}

{ While the phenomenon we want to investigate necessarily happens at finite temperature, we first need to study the system at the lowest temperatures in order to determine the magnetic-field region where it would be observable}. The unperturbed situation, where Eq.~\eqref{eq:dsim2} is limited to nearest-neighbour interactions ($D/|J_{\textit{eff}}|=0$), has been studied before in Ref.~\onlinecite{Guruciaga16}. In the absence of magnetic field, the ground state is a crystal of double charges alternating in sign, making a zinc blende structure with $\rho_S^d=1$~\cite{Guruciaga14}. Having no magnetic moment, double monopoles are unstable under any sufficiently strong magnetic field. Fig.~\ref{E-D_vs_B}(b) shows this for the relevant situation were the field is along the $[110]$ direction. We computed the change in energy $e(h)$ of a single tetrahedron in the nearest-neighbour approximation due to the increasing field $h$ in the Zeeman term. We classified spin configurations as neutral ($n$) sites, and double ($d$) and single ($s$) monopoles. Above an inferior field $h_{\textit{inf}}$ the $\alpha$-spin chains result completely polarised, and two particular single monopoles (one of each sign) are the stable configurations in each tetrahedron. Construction constraints {involving the underlying spins} impose antiferromagnetic order within $\beta$-spin chains, and they in turn form charge-ordered chains running perpendicular to the field (see Fig.~\ref{piroc}(c)). {The polarised $\alpha$ spins decouple each of these ordered $\beta$ chains from the rest. The ground state turns then into a $\rho^s = 1$ disordered array with subextensive residual entropy~\cite{Guruciaga16}, characterised by $\rho_S^d=0$ (there are no double charges) and $\rho_S^s=0$ }(with $D=0$ there is no driving force at $T=0$ to impose coherence between the charge-alternating single-monopole $\beta$ chains).

Although there are no further changes in the ground state on increasing field, it is important for the study of OBD to establish the field range where double monopoles are the lowest-lying energy excitations { out of the single-monopole ground state. It has been shown before that, due to construction constraints involving the spins that make the magnetic charges, these double monopoles favour pair monopole correlations even in the absence of dipolar interactions~\cite{Guruciaga14,Slobinsky2018charge}, while neutral sites disfavour them. Applied to the present situation, this general argument has important consequences. Imagine neutral sites are so energetically costly compared with double charges that their density is negligible at low temperatures. It is then easy to check that the only way to promote low energy excitations ---i.e., double monopoles--- is by turning an $\alpha$ spin linking plus and minus single charges in consecutive charge-ordered chains (see Fig.~\ref{piroc}(b) and (c)). As shown numerically and analytically in Ref.~\onlinecite{Guruciaga16}, the proliferation of these double charges at finite temperature} drives the appearance of long-range charge order by the coherent assembly of adjacent $\beta$ chains~\cite{Guruciaga16}. Fig.~\ref{E-D_vs_B}(b) allows us to extract the field range $[h_{\textit{inf}},h_{\textit{sup}}]$ (black dotted lines) where OBD was shown to be possible {for the particular case $D=0$}.

The inclusion of dipolar interactions ($D \neq 0$) is made conceptually and even quantitatively simple within the monopole picture. We expect the zinc-blende crystal of double charges to become more stable (through an extra term proportional to $-4DQ_s^{\, 2}$) and thus to extend to higher fields. In turn, the disordered ground state of single monopoles should now be ordered, stabilised by a term proportional to $-DQ_s^{\, 2}$. 
{This is why ---if dipolar energy was the main ordering force--- we would} expect the charge-ordered state of single monopoles to be destroyed at a critical temperature scaling with $D$ (i.e., comparatively small within the context of this study). {This {way of reasoning} is what will allow us to observe indirectly OBD effects even in the presence of such perturbations: we will choose to work at a ``high'' temperature $T \approx |J_{\textit{eff}}|/k_B$ such that the effect of perturbations is small, with OBD as the dominating driving force towards order. {As we will see in Sec.~\ref{Finite temperature}, the need to work at a relatively high temperature will make the upper limit $h_{\textit{sup}}$ much less significant than $h_{\textit{inf}}$.}}

In yet another example of its simplicity and beauty, we estimate now the field $h_{\textit{inf}}$ using the monopole picture. 
The Madelung energy per pair of particles of a crystal of magnetic charges $Q$ and $-Q$ in the zinc-blende structure is given by
\begin{equation}
    M=-\frac{\mu_0}{4\pi}\frac{\alpha\, Q^2}{r_{d}} \, ,
    \label{eq:madelung}
\end{equation}
with $\alpha = 1.6381$~\cite{Ashcroft1976solid} and $Q$ a function of $D$ through its link to the magnetic moment $\mu= \sqrt{({4\pi}/{\mu_0})r_{\textit{nn}}^3\, D}$. 
Their self-interaction energy~\cite{Castelnovo}, in turn, is 
\begin{equation}
    E^{{\textit{self}}}= 
    \left[J_{\textit{eff}}-\frac{5}{3}D+\frac{4}{5}\left(1+\sqrt{\frac{2}{3}}\right)\frac{5}{3}D\right] \left(\frac{r_d}{\mu}\,Q\right)^2 \, .
    \label{eq:self}
\end{equation}
The factor on the right is a constant taking the value $4$ or $8$ (depending on whether the crystal is made of single or double charges, respectively), rendering $E^{{\textit{self}}}$ a function of $D$ and $J_{\textit{eff}}$ only.
Taking this into account, $h_{\textit{inf}}$ can be calculated by simply considering the Zeeman energy of the configurations favoured by the field and solving
\begin{equation}
    E_d = E_s-\frac{4}{\sqrt 6}\, h_{\textit{inf}} \, ,
    \label{eq:Binf}
\end{equation}
{where the energy $E_s=M_s+E_s^{{\textit{self}}}$ related to the single-monopole crystal is equal to the sum of Eqs.~\eqref{eq:madelung} and~\eqref{eq:self} evaluated at $Q_s$. Similarly, $E_d=M_d+E_d^{{\textit{self}}}$ is the energy of the double monopole crystal}. 
We show the analytical predictions of the monopole picture for different values of $D$ in Fig.~\ref{E-D_vs_B}(a). We include also the results for $h_{\textit{inf}}$ and $h_{\textit{sup}}$ obtained by simulating small spin systems ($L=2$) at very low temperatures (down to $k_B T/|J_{\textit{eff}}|=0.18$)~\footnote{Although the lowest-possible temperatures are desirable in order to determine the ground-state transition field $h_{\textit{inf}}$, compromises have to be made not only to ensure equilibrium but also to resolve $h_{\textit{sup}}$, which involves excitations and is thus extremely subtle.} and extrapolating to $T=0$ the fields where the number densities of the different tetrahedron configurations of interest cross. { More explicitly, we looked for the condition $\rho^s=\rho^d$ to estimate $h_{\textit{inf}}$ (see Sec.~\ref{sec:NN}), and $\rho^n=\rho^d$ for $h_{\textit{sup}}$.\footnote{We will see in the next section that the density of double monopoles could be affected in a non-trivial way by very peculiar finite-size effects that induce disorder at very low temperatures for $D=0$~\cite{Guruciaga16}. However, the system remains ordered even at $T=0$ for $D\neq0$, which is the important case here.} Fig.~\ref{E-D_vs_B}(a) shows that, in spite of the approximations involved, both estimations for $h_{\textit{inf}}$ give very similar results.}


\subsection{Finite temperature} \label{Finite temperature}
\subsubsection{Nearest-neighbours interactions ($D=0$)}\label{sec:NN} 

The monopole picture implies that there is no energetic drive towards charge order for $D=0$. In spite of this, at low fields and temperatures we expect to find a crystal of double charges with the zinc-blende structure. {This ``all-spins-in--all-spins-out'' structure is stabilised not by charge-charge interactions, but due to the constraints imposed by the spins that make these charges}~\cite{Guruciaga14}. For $h \gtrsim h_{\textit{inf}}$ (which stabilises a disordered ground state of single monopoles) the situation differs from what it would be trivially expected. As first shown in Ref.~\onlinecite{Guruciaga16}, charge disorder should occur for $T$ \emph{strictly zero} as predicted but, as { we explained before and will see below, the proliferation of double charges favours an array of staggered magnetic charges at finite temperature}. { This array transitions to a disordered state only for $k_B T/|J_{\textit{eff}}| > 1$.}

\begin{figure}
    \centering
    \includegraphics[width=1\linewidth]{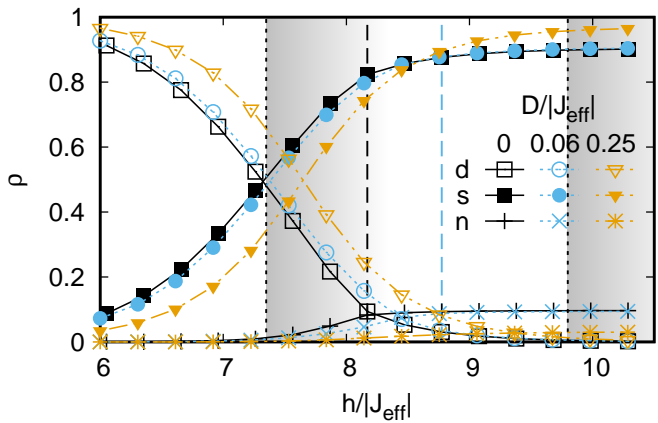}
    \caption{{Densities of single monopoles, double monopoles, and neutral sites vs. magnetic field at $k_BT/|J_{\textit{eff}}|=0.9$, and $L=6$. Colour identifies different values of the dipolar strength $D$ {(note that $|J_{\textit{eff}}|$ was kept constant throughout this work)}, while dashed vertical lines (with the same colour code) highlight the approximate field where charge disorder occurs for each value of $D$. Throughout the text, the colour black corresponds to the ideal case with $D=0$ and field exactly parallel to $[110]$ ($\theta=0$). The dotted vertical lines indicate the range of fields where the ground state consists of single charges, with double charges acting as the lowest-lying excitations. The horizontal grey gradient illustrates how the frontiers of this region moves to higher values of $h$ when $D>0$, as shown in Fig.~\ref{E-D_vs_B}(a). 
    As explained in the text, the transition field $h_{\textit{inf}}$ from a ground state of double charges to one of single charges can be approximately read from the point where $\rho^s$ and $\rho^d$ intersect.}}
    \label{rho-s-d-n_dip}
\end{figure}

\begin{figure}[t]
    \centering
    \includegraphics[width=1\linewidth]{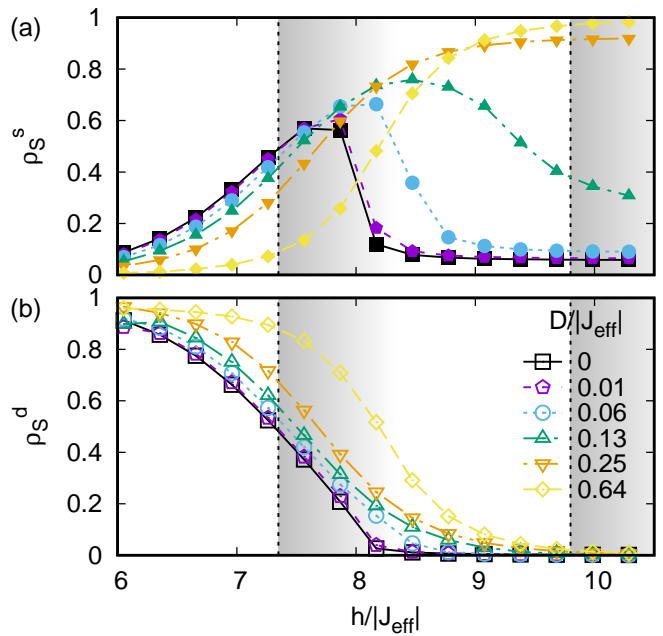}
    \caption{{(a)~Single ($\rho_S^s$) and (b)~double ($\rho_S^d$) monopole staggered densities as functions of the magnetic field for $k_B T/|J_{\textit{eff}}|=0.9$ and $L=6$. The colour coding marks the different values of the dipolar strength $D/|J_{\textit{eff}}|$. The dotted vertical lines highlight $h_{\textit{inf}}$ and $h_{\textit{sup}}$ for the ideal $D=0$ case, while the gradient represents that this region moves to higher values of $h$ when $D>0$. The shape of $\rho_S^s(h)$ changes drastically for $D > 0.13$, saturating at non-zero values at high fields even for big lattice sizes.}} 
    \label{rhoSs-d_vs_B}
\end{figure}

\begin{figure}[t]
    \centering
    \includegraphics[width=1\linewidth]{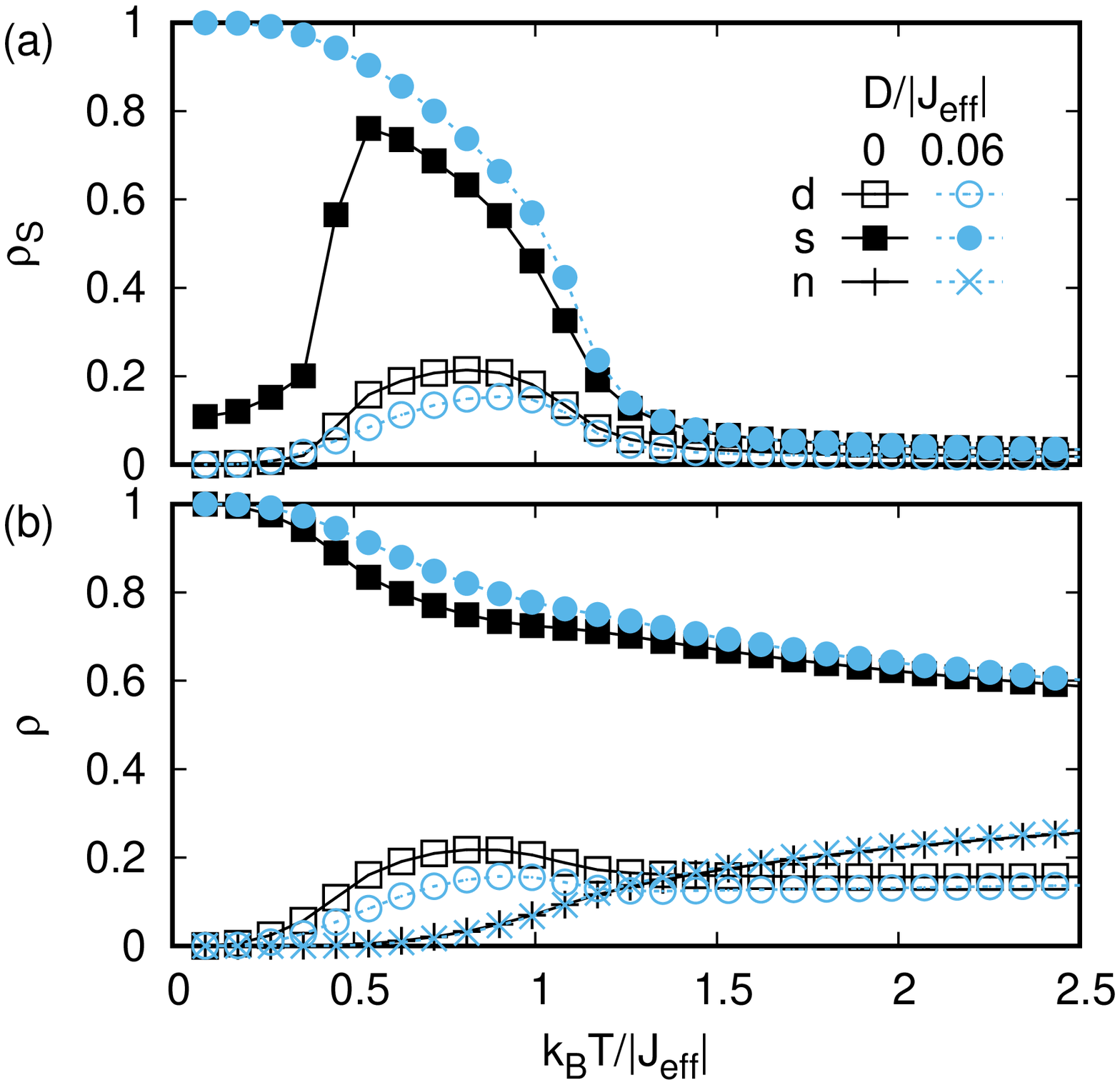}
    \caption{(a)~{Single and double-monopole staggered densities vs. temperature. (b)~Number densities of single, double charges, and neutral sites vs. temperature ($L=6$). The curves were measured for the ideal $D=0$ case and $D/|J_{\textit{eff}}|=0.06$. The magnetic field in the $[110]$ direction was $h/|J_{\textit{eff}}|=7.86$ and $8.16$, respectively, where the corresponding $\rho_S^s$ is maximum for $k_BT/|J_{\textit{eff}}|=0.9$ (see Fig.~\ref{rhoSs-d_vs_B}(a)). In close relation with the OBD effect, the density of double monopoles has a non-monotonic behaviour as a function of $T$.}}
    \label{rhoSs-d_vs_T}
\end{figure}

Figs.~\ref{rho-s-d-n_dip},~\ref{rhoSs-d_vs_B}, and~\ref{rhoSs-d_vs_T} show the results of Monte Carlo simulations for a field parallel to $[110]$ and $D=0$ (black symbols and lines), in a system with $L=6$. We can see in Fig.~\ref{rho-s-d-n_dip} that the number density of the different {monopoles} in a tetrahedron evolves with increasing magnetic field at $k_B T/|J_{\textit{eff}}| = 0.9$ from a majority of double monopoles to one of single charges, as expected. At the highest fields, neutral sites are the main excitation. {Due to the energy scales involved we expect to have at these temperatures only: \textit{i-} positive and negative double monopoles, \textit{ii-} the two types of single monopoles with magnetic moment direction favoured by the magnetic field, and \textit{iii-} the same for neutral sites (see continuous lines in Fig.~\ref{E-D_vs_B}(b)). This considerably helps the analysis; for example, to understand that at the transition field $h_{\textit{inf}}$ the density of double and single charges is $\rho^d\approx \rho^s\approx 0.5$ (curves crossing in Fig.~\ref{rho-s-d-n_dip}). In the same terms, it may be expected that the curves for the double-monopole and neutral-sites densities cross very near $h_{\textit{sup}}$. However, against intuition, we can see them in this last case crossing at much smaller fields than that. While we will be able to explain this when analysing the order parameter behaviour, we can advance now that this rapid proliferation of neutral sites at low fields implies the destruction of OBD at $h$ much smaller than $h_{\textit{sup}}$.}

The relevant order parameters (the staggered charge density of double and single monopoles) are plotted in Figs.~\ref{rhoSs-d_vs_B}(a) and~\ref{rhoSs-d_vs_B}(b) as functions of field at the same working temperature $k_B T/|J_{\textit{eff}}| = 0.9$. The relatively big value of $\rho_S^d$ in Fig.~\ref{rhoSs-d_vs_B}(b) at low fields reflects the zinc-blende structure of double monopoles, gradually fading with increasing $h$. The fact that $\rho_S^d \neq 0$ for $h\gtrsim h_{\textit{inf}}$ reveals that a considerable number of ordered double monopoles {(created by thermal fluctuations)} remain even above this threshold.
The persistence of double-monopole order above $h_{\textit{inf}}$ and the continuous aspect of the curves on traversing the field threshold makes it tempting to think that these measurements ---performed on increasing field--- are affected by the past history, retaining at high fields some of the zinc-blende structure which characterises order at lower ones. This is not the case: the same results are obtained in field-decreasing runs at fixed $T$ (not shown), and at fixed field and decreasing temperature (Fig.~\ref{rhoSs-d_vs_T}(a)). {As mentioned before and made clear in this same figure, although the system should be disordered at zero temperature, charge order exists at finite temperature below the scale of $|J_{\textit{eff}}|/k_B$.}

The behaviour of $\rho_S^s$ seems to be complementary to that of $\rho_S^d$, with an increasing amount of single monopole order at fields above $h_{\textit{inf}}$.
The establishment of staggered single-charge order at this finite $T$ above $h_{\textit{inf}}$ has been analytically and numerically studied as a textbook case of field-tuned OBD, triggered by a diverging peak in the density of states (see Fig. 3 in Ref.~\onlinecite{Guruciaga16}) associated with this particular monopole array. 
Regarding the single charge order at fields below $h_{\textit{inf}}$, it can be naturally interpreted in terms of the underlying ``all-in--all-out'' double-monopole crystal lattice, to which single charges are the lowest energy excitations (Fig.~\ref{E-D_vs_B}(b)). The converse is true for the observed finite value of $\rho_S^d$ for $h \gtrsim h_{\textit{inf}}$.

{In relation to this, we owe an explanation for the reduced value of $\rho^d$ with respect to $\rho^n$ at moderate fields $h \lesssim h_{\textit{sup}}$, observed in Fig.~\ref{rho-s-d-n_dip}. At very low temperatures neutral sites are almost banned by their high energy cost, while the excitation of the allowed quasiparticles (double monopoles) imply the existence of staggered single-charge order~\cite{Guruciaga16}. At higher temperatures, the arousal of neutral charges help to decorrelate the $\beta$ chains, and the staggered order is lost (Fig.~\ref{rhoSs-d_vs_T}(a)). The absence of order among consecutive $\beta$ chains makes now less probable to excite double monopoles (which are mainly created by flipping $\alpha$ spins that link staggered-ordered $\beta$ chains (see Fig.~\ref{piroc}(c) and Ref.~\onlinecite{Guruciaga16})). Since neutral sites are not so constrained {(i.e., they do not need the coherent order of consecutive $\beta$ chains to appear)} this explains the crossing of their density curves at finite $T$ below $h_{\textit{sup}}$ observed in Fig.~\ref{rho-s-d-n_dip}(a). To illustrate this phenomenon, Fig.~\ref{rhoSs-d_vs_T}(b) shows the density of monopoles, now at fixed $h$ and as a function of temperature (black lines and symbols). We would naively have expected a density of double monopoles monotonously increasing with temperature. We can see that while $\rho_S^s$ decreases near the transition at $k_B T \approx |J_{\textit{eff}}|$ (Fig.~\ref{rhoSs-d_vs_T}(a)) the number of double monopoles starts to decrease, producing a maximum in its density within the staggered-order region.} {This finite temperature phenomenon implies no contradiction with the method we used to find $h_{\textit{sup}}$, performed in the limit of zero temperature}.

The abrupt fall on $\rho_S^s(h)$ at fixed temperature on increasing fields observed in Fig.~\ref{rhoSs-d_vs_B}(a) is one of the main results we have to show in this work. It relates to the switching-off of the entropic drive towards order (the only one present in the current situation) produced by the proliferation of neutral sites~\cite{Guruciaga14,Slobinsky2018charge}. A key idea is that both the thermally-induced {single-charge} correlations produced by double monopoles at low fields, and the decorrelating effect of neutral sites that causes a drop in $\rho_S^s$ at higher ones, will still be operative if dipolar interactions or other perturbations are present. As will be shown in the next subsection, this fact 
can be used to mark the presence of OBD effects.

\subsubsection{Dipolar interactions (finite $D$)}

Figs.~\ref{rhoSs-d_vs_B}(a) and~\ref{rhoSs-d_vs_B}(b) present $\rho_S^s$ and $\rho_S^d$ as functions of the field when long-range dipolar interactions are included. These curves were measured in the same conditions as the nearest-neighbour case. It can be seen that as $D$ grows, an increasing field $h$ is needed to destabilise the staggered order of double monopoles in favour of an order of single monopoles, as predicted by Fig.~\ref{E-D_vs_B}(a). The shape of the curves $\rho_S^s(h)$ allows us to separate them (at this particular working temperature) in two distinct sets: values of $D$ smaller and bigger than $D/|J_{\textit{eff}}| = 0.13$.

\begin{figure}[t]
    \centering
    \includegraphics[width=1\linewidth]{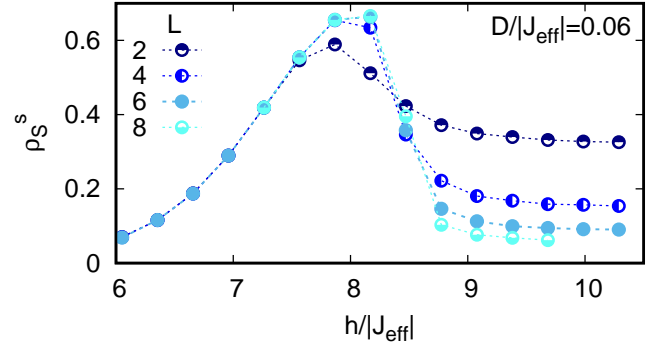}
    \caption{{Evolution of the single-monopole staggered density with system size for $D/|J_{\textit{eff}}|=0.06$ at $k_BT/|J_{\textit{eff}}|=0.9$. The behaviour of the order parameter persists with increasing system size.}}
    \label{finsize}
\end{figure}

The group of smaller $D$ is the relevant one concerning the main objective of this paper. {Within this set, the temperature $T=0.9|J_{\textit{eff}}|/k_B$ is so high that Coulomb forces among monopoles cannot hold the staggered charge array. The curves resemble then} those obtained without dipolar long-range interactions, showing that the main effect of a finite $D$ is to shift $h_{{\textit{inf}}}$, but preserving the same physics at this temperature. The abrupt fall of the order parameter down to $\rho_S^s=0$ on increasing $h$, {which persists in the thermodynamic limit (Fig.~\ref{finsize})}, mirrors the switching-off of the OBD mechanism and serves as an indication of its relevance in the stabilisation of order.

{A non-trivial behaviour} associated with OBD is exemplified by Fig.~\ref{rhoSs-d_vs_T}(a). There, the order parameter $\rho_S^s$ is compared for $D=0$ (black curves) and $D/|J_{\textit{eff}}|=0.06$ (blue), now as a function of temperature at $h/|J_{\textit{eff}}|=7.86$ and $8.16$, respectively, where the corresponding $\rho_S^s(h)$ in Fig.~\ref{rhoSs-d_vs_B}(a) has a maximum. Diverting from the blue curve, $\rho_S^s$ for $D=0$ tends to zero at very low temperatures (and, evidently, for a constant lattice size $L$). This has been shown in Ref.~\onlinecite{Guruciaga16} to be a very peculiar finite-size effect characterising this system {(on increasing the size of the lattice the ordering trend continues on decreasing temperature, with disorder prevailing only at $T=0$)} that could be used as a method to identify OBD in artificial two-dimensional materials. On the other hand, $\rho_S^s$ continues growing when the system is cooled down at finite $D$: on reducing thermal fluctuations such that $k_B T/D \to 0$, dipolar interactions are able to promote monopole order. Regarding this usual measure of the relative value of thermal fluctuations compared with the tendency towards order (here, $k_B T/D$), there is a further aspect to note. Near the order-disorder transition ($k_B T/|J_{\textit{eff}}| \approx 1$ for both cases in Fig.~\ref{rhoSs-d_vs_T}(a)) this measure is quite big, $k_B T/D \approx 20$ for $D/|J_{\textit{eff}}|=0.06$ (of the order of $100$ for $D/|J_{\textit{eff}}|=0.01$), while it diverges for $D=0$. This makes us wonder once again about how much against intuition it is that charge order could be sustained {in the ideal case} in spite of the total absence of charge attraction and repulsion.

For the second group ($D/|J_{\textit{eff}}| \gtrsim 0.13$) the shape of the curves in Fig.~\ref{rhoSs-d_vs_B}(a) starts to distort considerably. Their most salient feature is that there is no drop in the order parameter $\rho_S^s$ at high fields (where the corresponding single-charge densities are essentially saturated, as observed for the orange curve in Fig.~\ref{rho-s-d-n_dip}). This indicates that the Coulomb forces can now maintain the order of charges at this temperature. It can also be noted that {$\rho^s \approx 1$ at high fields: the concentration} of doubles is made unstable by the Zeeman energy, while neutral sites are disfavoured by the Coulomb term hidden in Eq.~\eqref{eq:dsim2}. 

\begin{figure*}[t]
    \centering
    \includegraphics[width=\linewidth]{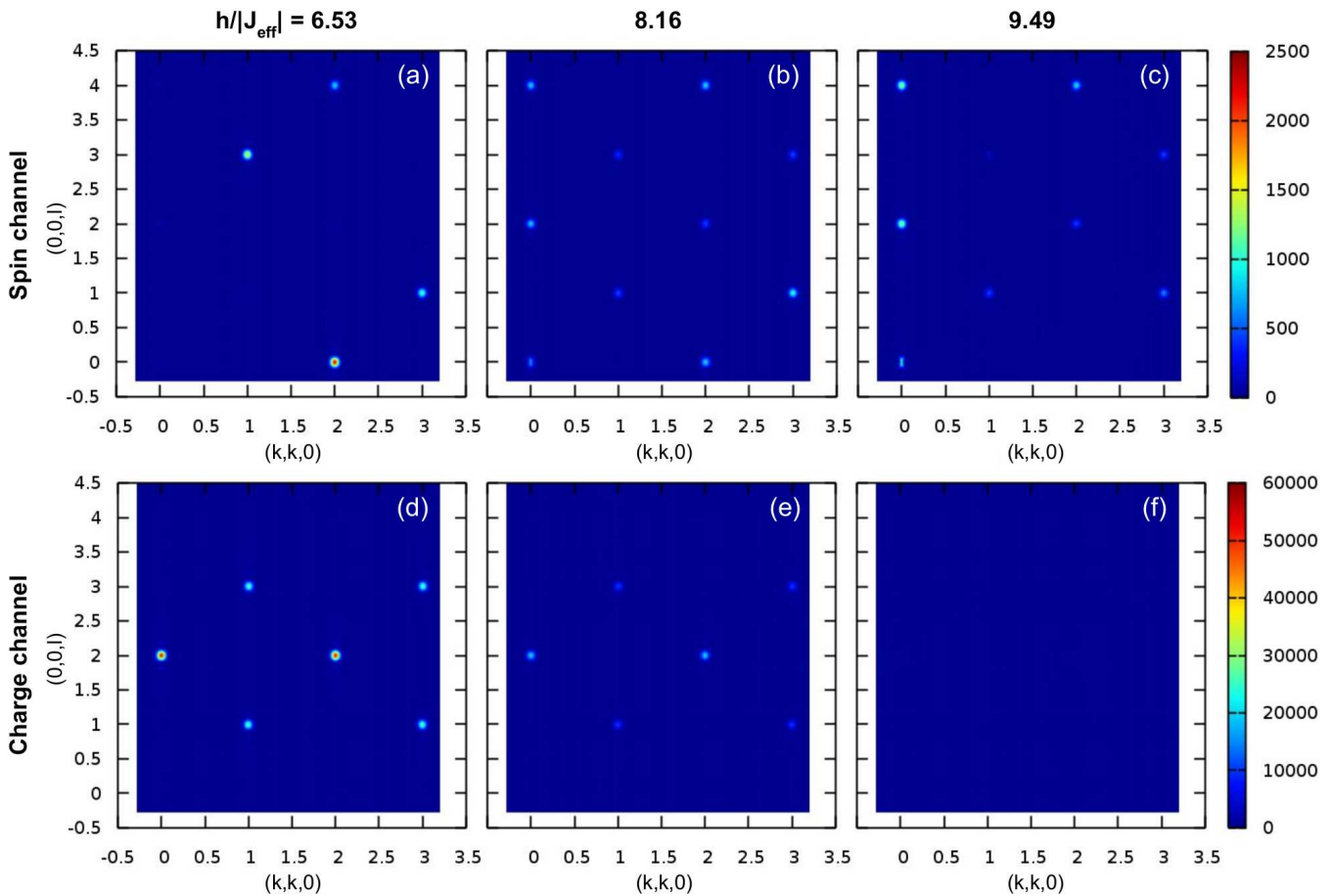}
    \caption{{(a)-(c)~Spin structure factor and (d)-(f)~charge-charge correlation function in the $(kkl)$ plane. The different values of the field $h$ in the $[110]$ direction characterise the three types of order: crystal of double monopoles, crystal of single monopoles, and disorder. In all cases $D/|J_{\textit{eff}}|=0.06$, $k_BT/|J_{\textit{eff}}|=0.9$ and $L=8$. We can see that the intense Bragg peak at $(220)$ first diminishes its intensity and then disappears on increasing field (panels~(a),~(b) and~(c)), while the double-charge order gives way to a single-charge crystal, and then to a disordered-charge phase with no Bragg peaks (panels~(d),~(e) and~(f)).}}
    \label{Sk}
\end{figure*}

Being $\rho_S^s$ and $\rho_S^d$ quantities generally inaccessible to most {laboratory-based experiments}~\footnote{The measurement of magnetocapacitance may be an interesting exception to this statement, {allowing an indirect approach to the formation of staggered monopole order} (see Refs.~\onlinecite{Katsufuji2004magnetocapacitance,Saito2005magnetodielectric,Khomskii2012electric}).}, an interesting probe to search for classical OBD in these magnetic systems would be neutron scattering. Figs.~\ref{Sk}(a)-(c) show the simulated spin structure factor (Eq.~\eqref{spinSk}) for a system with $D/|J_{\textit{eff}}|=0.06$ and $L=8$ at the working temperature $k_BT/|J_{\textit{eff}}|=0.9$. {We have chosen three values of the field, representative of the three types of order (or lack thereof) in the system.}
The patterns reflect the changes in magnetic order as the field turns the ``all-in--all-out'' structure (Fig.~\ref{Sk}(a)) into a polarised crystal of single monopoles (Fig.~\ref{Sk}(b)), and then into a disordered array of ordered $\beta$-spin chains (Fig.~\ref{Sk}(c)). The evolution of the $(220)$ peak can be used to give a quantitative measure of three-dimensional order (i.e. between different $\beta$ chains).
Supporting this, Figs.~\ref{Sk}(d)-(f) show the charge-charge correlation function defined in Eq.~\eqref{chargeSk} (which cannot be directly measured using neutrons, but illustrates well the parallel situation in the monopole framework). We can see that the sharp Bragg peaks associated with a crystal of monopoles change their intensity when passing from double to single charge order, and finally vanish for $h\approx h_{\textit{sup}}$, where the proliferation of neutral excitations promotes the disordering of the system.

{There exist many ``all-in--all-out'' materials in nature, some of which could be used for an experimental contrast against these ideas~\cite{Sleight1974semiconductor,Lhotel2015fluctuations,Xu2015,Anand15,Bertin2015nd}. However, it is important to bear in mind that each compound has its singularities, differing from the ideal case we study in its own way. \chem{Nd_2Hf_2O_7} and \chem{Nd_2Zr_2O_7}, for example, are pyrochlores with a big Ising anisotropy. They have relatively small N\'eel temperatures ($T_N \lesssim 0.5\text{ K}$) and big magnetic moments on the \chem{Nd} ion ($\approx 2.5\, \mu_B$), reduced by quantum fluctuations~\cite{Lhotel2015fluctuations,Anand15,Xu2015,Anand2017muon}. While \chem{Nd_2Zr_2O_7} is a good candidate to test the behaviour of our system on the high $D/|J_{\textit{eff}}|$ side~\cite{Xu2015}, \chem{Nd_2Hf_2O_7} has $D/|J_{\textit{eff}}|\approx 0.11$~\cite{Anand15}, slightly into the region where a decrease in order with increasing field would be expected.
Finding an ``all-in--all-out'' compound with \emph{very} small $D/|J_{\textit{eff}}|$ ratio to conclusively detect thermal OBD is, to our knowledge, much harder. The compound \chem{Cd_2Os_2O_7}, with quite a big ordering temperature (above $220\text{ K}$)~\cite{Yamaura2012phase}, could be a starting point, although it is clear that the physics implied (which has been studied since 1974~\cite{Sleight1974semiconductor}) transcends that of a simple system with localised Ising moments.}


\section{Effect of field misalignment}

While it is pretty easy in a numerical simulation, applying a field perfectly oriented in a particular crystallographic axis is something practically impossible in the experimental realm. This fact strongly impacts on the possibility of detecting the OBD phenomenon in the way we propose here.
A deviation of the field within the $x-y$ plane facilitates the formation of neutral tetrahedra without breaking the symmetry between the two FCC sublattices making the diamond structure. Therefore, it preserves spontaneous OBD but lowers the disordering temperature. On the other hand, a nonzero component of $\mathbf B$ in the $[001]$ direction explicitly breaks the relevant symmetry, and thus favours the occurrence of single-charge order. It is easy to understand this by noting that now the field has a larger component parallel to $[111]$, which stabilises the order of positive single charges in up tetrahedra
and negative in down tetrahedra.~\footnotetext[100]{The $[110]$ direction has the special feature of being just midway between $[111]$ and $[11\overline{1}]$; the latter promotes the opposite order.}~\cite{Note100,Sakakibara03}

\begin{figure}
    \centering
    \includegraphics[width=1\linewidth]{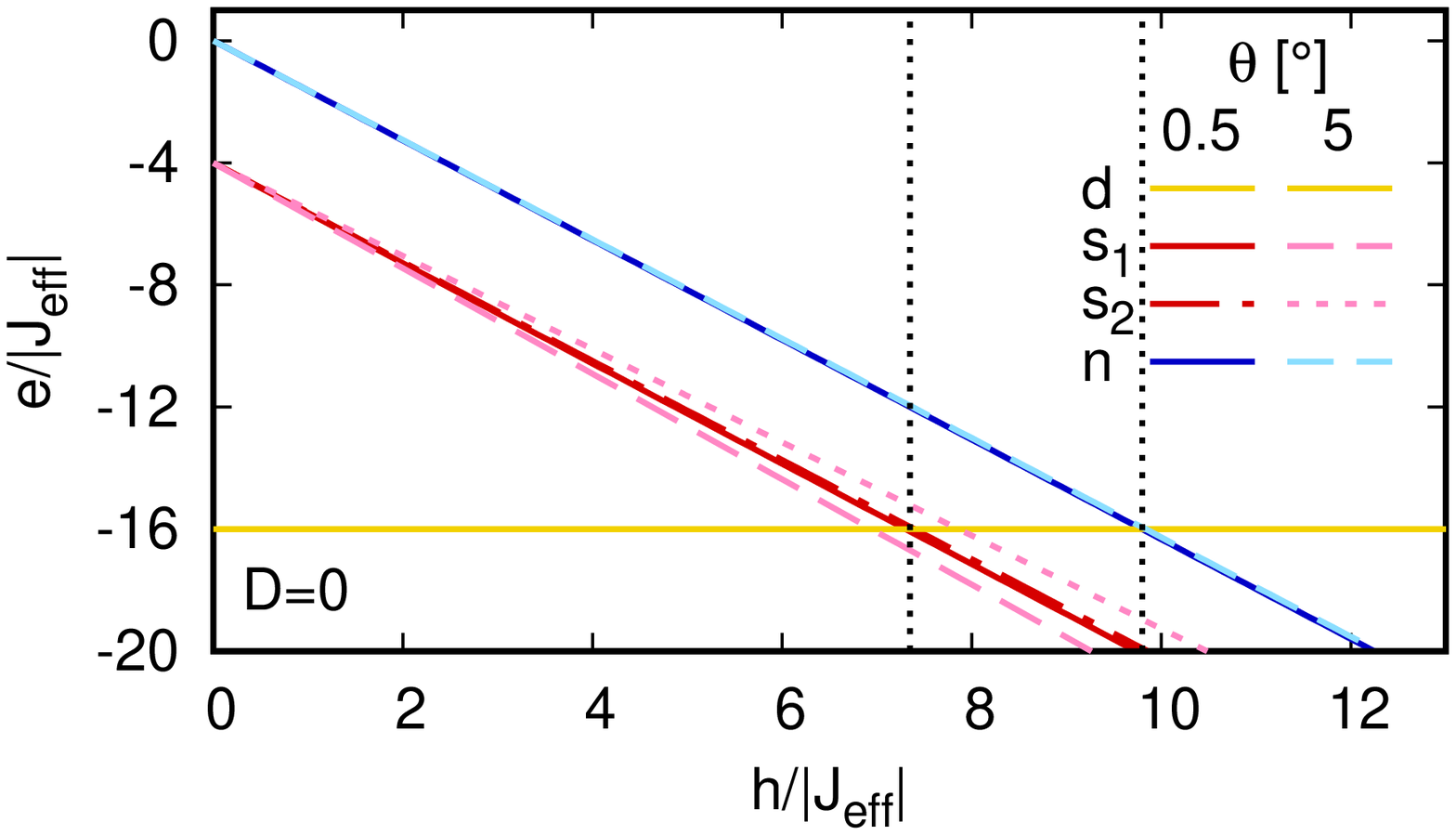}
    \caption{{Energy vs. field for the different low-lying spin configurations in {an up} tetrahedron with nearest-neighbours interactions ($D=0$). The magnetic field vector $\mathbf{B}$ of intensity $h/\mu$ is tilted anticlockwise an angle $\theta$ from the $[110]$ direction, around $\hat{\mathbf{c}}\propto [1\overline{1}0]$ (see Fig.~\ref{piroc}). The introduction of a nonzero component along $[001]$ produces a splitting of the single-charge energy levels, but the $s_1$ and $s_2$ curves are effectively indistinguishable in the temperature and field range of interest for $\theta < 1^{\circ}$. The dotted vertical lines highlight the region where the ground state is composed of single charges with double charges as the low energy excitations, for the ideal $\theta=0$ case.}}
    \label{E_vs_B_tilt}
\end{figure}

As mentioned in the introduction, this case is different from other perturbation terms, since the perturbation coupled with the order parameter grows in parallel with the magnetic field $h$, which is the instrument we use to detect the entropic drive towards order. To assess the significance of this effect at finite $T$, we consider the nearest-neighbour model (Eq.~\eqref{eq:dsim} with $D=0$) under a magnetic field $\mathbf{B}$ of strength $h/\mu$ tilted an angle $\theta$ from the $[110]$ direction towards $[001]$ (i.e., $\theta$ represents an anticlockwise rotation of that magnitude around $\hat{\mathbf{c}}\propto[1\overline{1}0] = [110]\times [001]$, see Figs.~\ref{piroc}(a) and~\ref{piroc}(b)). 
As shown in Fig.~\ref{E_vs_B_tilt}, the presence of $h_{001}=h\sin\theta\neq 0$ splits the low-lying single-charge energy levels {associated to each type of tetrahedron (up or down)}, opening a gap proportional to $h_{001}$. For any value of $\theta\neq0$, the single-charge ground state is strictly composed of one particular configuration and thus ordered. However, for small-enough values of $\theta$, the gap is effectively nonexistent in the range of $h$ and $T$ in which we are interested throughout this work, and the misaligned system can still show a signature of OBD. 

\begin{figure}
    \centering
    \includegraphics[width=1\linewidth]{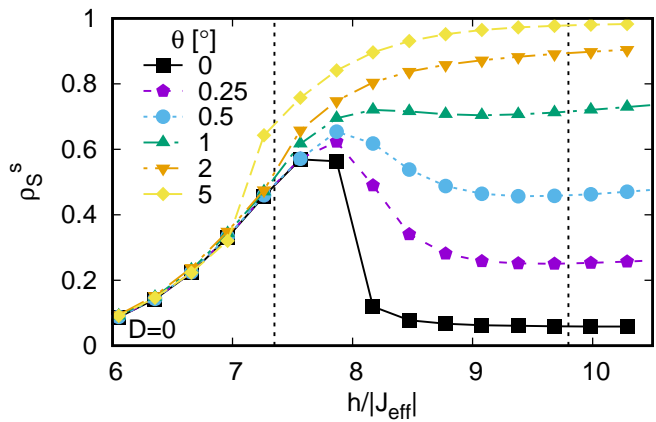}
    \caption{{Single-monopole staggered density vs. magnetic field for $k_BT/|J_{\textit{eff}}|=0.9$, $D=0$, and $L=6$. The colours refer to the diverse values of the small tilting angle $\theta$. Aside from $\theta=0$ (black curve) and differently from the case of dipolar interactions, any finite perturbation (i.e., any finite tilting angle) leads to a nonzero order parameter $\rho_S^s$ at high fields, even in the infinite size limit. Although a nonzero component of the field in the $[001]$ direction explicitly favours the occurrence of single-charge order, it is still possible to detect OBD for values of $\theta<1^{\circ}$, associated to the decrease in $\rho_S^s$.
    The vertical lines correspond to the ideal case ($\theta=0$).}}
    \label{rhoSs-vs-H_tilt}
\end{figure}

Fig.~\ref{rhoSs-vs-H_tilt} shows the relevant order parameter $\rho_S^s$ as a function of the modulus of the field, for different values of the tilting angle $\theta$. This figure is analogous to Fig.~\ref{rhoSs-d_vs_B}(a). It was measured at the same temperature and for the same system size, but with a different perturbation term. In black, we include as a reference the curve with a perfectly aligned field; given that $D=0$, it is the same black curve as the one in Fig.~\ref{rhoSs-d_vs_B}(a). The general shapes of the curves are similar to those encountered before. However, for all finite values of $\theta$, we observe at high $h$ a finite (and indeed, slightly increasing with field) value of $\rho_S^s$. As cautioned before, this marks an important difference with the previous perturbation: the system remains charge-ordered at high fields in any real situation. 

Striking the main target we have fixed for this paper, the sudden drop observed in $\rho_S^s$ at very small angles (well below $1^{\circ}$) can only be explained by the weakening of the OBD effect that holds the order of single monopoles at intermediate fields even for $\theta=0$. Again, the observation of this drop is an indirect proof of the OBD mechanism being operative. The need of such an accuracy in the field angle control certainly complicates the observation of the effect, but it should not be considered as an experimental impediment: triple-axis vector magnets (as, for example, in Ref.~\onlinecite{Bruin2013}), {high-precision goniometers (like in Ref.~\onlinecite{Sazonov2010field}) or in-situ sample rotators~\cite{Kenzelmann2019} can be used to correct misalignments below the needed accuracy in a neutron scattering experiment.}


\section{Discussion}

As advertised before, in our system the dipolar perturbation term that favours order remains constant when $h$ increases; this allows to recognise the effect in the order parameter of the field-tuned entropic contribution. However, it may be important to consider that bigger fields favour the introduction of neutral sites at the expense of double monopoles, not only weakening the thermal drive towards order, but also slightly diminishing the effectiveness of the Coulomb interactions. For small values of $D$ ($D/|J_{\textit{eff}}| < 0.03$) the differences observed in monopole densities with respect to $D=0$ (not shown) are negligible, making evident that the OBD effect is the main driving force. {For $D/|J_{\textit{eff}}| = 0.06$, the field at which charge disorder occurs is somewhat bigger to that of $D=0$ (dashed lines in Fig.~\ref{rho-s-d-n_dip}). Although the density of neutral sites at the transition fields is quite comparable, the density of double monopoles for $D/|J_{\textit{eff}}| = 0.06$ is about half of the value it takes at the transition for $D=0$. This signals that, although dipolar interactions are not enough to hold the single-monopole crystal together at high fields, they are helping to sustain it near the order-disorder transition.}

{In a final note we will refer to the energy scales of the relevant perturbations needed to sustain order, measured with respect to the thermal energy. Naively speaking, it may result quite surprising that angular deviations from perfect alignment as small as those mentioned before can have such big effects. Take $\theta=0.25^{\circ}$ as an example: how can a perturbation of $h_{001}= 9|J_{\textit{eff}}|\,\sin(0.25^{\circ})\approx 0.05\,k_B T$ create an order parameter as big as $\rho_S^s\approx 0.25$ for $h/|J_{\textit{eff}}|=9$?
{We think that the answer to this question relies on the fact that this is not the only energy scale in the system. Indeed, $J_{\textit{eff}}$ is the most relevant energy value here. It guarantees that partial single-charge order is retained within $\beta$ chains at this $T$, so that spins do not respond to the misaligned magnetic field on their own, but as part of finite $\beta$ chains (see, for example, Ref.~\onlinecite{Sazonov2012double}).} Something similar happens when dipolar interactions act as a perturbation, explaining that Coulomb forces can sustain single-charge order at big fields when $D/|J_{\textit{eff}}| \approx D/k_B T \approx 0.13$.}


\section{Conclusions}
Order by thermal disorder is a very peculiar effect that has been studied for almost forty years in the context of frustrated magnetic systems. In spite of this, it has associated very few validated examples in nature. In this work we proposed a scenario (the Ising pyrochlore with antiferromagnetic ``all-in--all-out'' order in zero field) where it may be detected. An easily-controllable external parameter (a magnetic field in the $[110]$ crystallographic direction) can be used to turn on and off the entropic drive towards order, unmasking thus the effect even in the presence of perturbations coupling to the order parameter.
Following a protocol that in principle can be applied to any other uncontrolled energy interactions, we showed that OBD can still be found when dipolar interactions are present, provided that they are weak with respect to the nearest-neighbour exchange interaction ($D<0.13\,|J_{\textit{eff}}|$ for a working temperature $T\approx |J_{\textit{eff}}|/k_B$). 
However, special care needs to be taken in order to ensure a field-misalignment angle from the $[110]$ direction well below $1^{\circ}$. This could be attainable in neutron scattering measurements, {one of the techniques} we propose to identify order due to thermal fluctuations.
Although we know of no ideal material where these experiment could be performed, the measurement of a fast decrease in intensity of the $(220)$ peak in the spin structure factor with increasing field in ``all-in--all-out'' materials like \chem{Nd_2Hf_2O_7} or \chem{Cd_2Os_2O_7} would be a clear signature of the OBD mechanism being operative.


\begin{acknowledgments}
We thank H.~D.~Rosales for help with the structure factor code, M.~J.~P.~Gingras for useful input on the relevant literature, G.~Aurelio and M.~Kenzelmann for insights about angle control in neutron scattering experiments, and UnCaFiQT (SNCAD) for providing computational resources. The authors acknowledge financial support from Agencia Nacional de Promoci\'on Cient\'\i fica y Tecnol\'ogica (ANPCyT) through grants PICT 2013 N$^{\circ}$2004 and PICT 2014 N$^{\circ}$2618 and Consejo Nacional de Investigaciones Cient\'\i ficas y T\'ecnicas (CONICET) through grant PIP 0446.
\end{acknowledgments}



\begin{thebibliography}{59}%
\makeatletter
\providecommand \@ifxundefined [1]{%
 \@ifx{#1\undefined}
}%
\providecommand \@ifnum [1]{%
 \ifnum #1\expandafter \@firstoftwo
 \else \expandafter \@secondoftwo
 \fi
}%
\providecommand \@ifx [1]{%
 \ifx #1\expandafter \@firstoftwo
 \else \expandafter \@secondoftwo
 \fi
}%
\providecommand \natexlab [1]{#1}%
\providecommand \enquote  [1]{``#1''}%
\providecommand \bibnamefont  [1]{#1}%
\providecommand \bibfnamefont [1]{#1}%
\providecommand \citenamefont [1]{#1}%
\providecommand \href@noop [0]{\@secondoftwo}%
\providecommand \href [0]{\begingroup \@sanitize@url \@href}%
\providecommand \@href[1]{\@@startlink{#1}\@@href}%
\providecommand \@@href[1]{\endgroup#1\@@endlink}%
\providecommand \@sanitize@url [0]{\catcode `\\12\catcode `\$12\catcode
  `\&12\catcode `\#12\catcode `\^12\catcode `\_12\catcode `\%12\relax}%
\providecommand \@@startlink[1]{}%
\providecommand \@@endlink[0]{}%
\providecommand \url  [0]{\begingroup\@sanitize@url \@url }%
\providecommand \@url [1]{\endgroup\@href {#1}{\urlprefix }}%
\providecommand \urlprefix  [0]{URL }%
\providecommand \Eprint [0]{\href }%
\providecommand \doibase [0]{http://dx.doi.org/}%
\providecommand \selectlanguage [0]{\@gobble}%
\providecommand \bibinfo  [0]{\@secondoftwo}%
\providecommand \bibfield  [0]{\@secondoftwo}%
\providecommand \translation [1]{[#1]}%
\providecommand \BibitemOpen [0]{}%
\providecommand \bibitemStop [0]{}%
\providecommand \bibitemNoStop [0]{.\EOS\space}%
\providecommand \EOS [0]{\spacefactor3000\relax}%
\providecommand \BibitemShut  [1]{\csname bibitem#1\endcsname}%
\let\auto@bib@innerbib\@empty
\bibitem [{\citenamefont {Cardy}(1996)}]{Cardy1996scaling}%
  \BibitemOpen
  \bibfield  {author} {\bibinfo {author} {\bibfnamefont {J.}~\bibnamefont
  {Cardy}},\ }\href@noop {} {\emph {\bibinfo {title} {Scaling and
  Renormalization in Statistical Physics}}},\ Vol.~\bibinfo {volume} {5}\
  (\bibinfo  {publisher} {Cambridge University Press},\ \bibinfo {year}
  {1996})\BibitemShut {NoStop}%
\bibitem [{\citenamefont {Ramirez}(1994)}]{Ramirez1994review}%
  \BibitemOpen
  \bibfield  {author} {\bibinfo {author} {\bibfnamefont {A.~P.}\ \bibnamefont
  {Ramirez}},\ }\href
  {https://www.annualreviews.org/doi/10.1146/annurev.ms.24.080194.002321}
  {\bibfield  {journal} {\bibinfo  {journal} {Annu. Rev. Mater. Sci.}\ }\textbf
  {\bibinfo {volume} {24}},\ \bibinfo {pages} {453} (\bibinfo {year}
  {1994})}\BibitemShut {NoStop}%
\bibitem [{\citenamefont {Moessner}\ and\ \citenamefont
  {Ramirez}(2006)}]{Moessner2006review}%
  \BibitemOpen
  \bibfield  {author} {\bibinfo {author} {\bibfnamefont {R.}~\bibnamefont
  {Moessner}}\ and\ \bibinfo {author} {\bibfnamefont {A.~P.}\ \bibnamefont
  {Ramirez}},\ }\href
  {https://physicstoday.scitation.org/doi/10.1063/1.2186278} {\bibfield
  {journal} {\bibinfo  {journal} {Physics Today}\ }\textbf {\bibinfo {volume}
  {59}},\ \bibinfo {pages} {24} (\bibinfo {year} {2006})}\BibitemShut {NoStop}%
\bibitem [{\citenamefont {Balents}(2010)}]{Balents2010spin}%
  \BibitemOpen
  \bibfield  {author} {\bibinfo {author} {\bibfnamefont {L.}~\bibnamefont
  {Balents}},\ }\href {https://www.nature.com/articles/nature08917} {\bibfield
  {journal} {\bibinfo  {journal} {Nature}\ }\textbf {\bibinfo {volume} {464}},\
  \bibinfo {pages} {199} (\bibinfo {year} {2010})}\BibitemShut {NoStop}%
\bibitem [{\citenamefont {Lacroix}\ \emph {et~al.}(2011)\citenamefont
  {Lacroix}, \citenamefont {Mendels},\ and\ \citenamefont
  {Mila}}]{Lacroix2011introduction}%
  \BibitemOpen
  \bibfield  {author} {\bibinfo {author} {\bibfnamefont {C.}~\bibnamefont
  {Lacroix}}, \bibinfo {author} {\bibfnamefont {P.}~\bibnamefont {Mendels}}, \
  and\ \bibinfo {author} {\bibfnamefont {F.}~\bibnamefont {Mila}},\ }\href@noop
  {} {\emph {\bibinfo {title} {Introduction to frustrated magnetism: materials,
  experiments, theory}}},\ Vol.\ \bibinfo {volume} {164}\ (\bibinfo
  {publisher} {Springer Science \& Business Media},\ \bibinfo {year}
  {2011})\BibitemShut {NoStop}%
\bibitem [{\citenamefont {Bramwell}\ and\ \citenamefont
  {Gingras}(2001)}]{Bramwell01}%
  \BibitemOpen
  \bibfield  {author} {\bibinfo {author} {\bibfnamefont {S.~T.}\ \bibnamefont
  {Bramwell}}\ and\ \bibinfo {author} {\bibfnamefont {M.~J.~P.}\ \bibnamefont
  {Gingras}},\ }\href {\doibase 10.1126/science.1064761} {\bibfield  {journal}
  {\bibinfo  {journal} {Science}\ }\textbf {\bibinfo {volume} {294}},\ \bibinfo
  {pages} {1495} (\bibinfo {year} {2001})}\BibitemShut {NoStop}%
\bibitem [{\citenamefont {Diep}(2013)}]{Diep2013frustrated}%
  \BibitemOpen
  \bibfield  {author} {\bibinfo {author} {\bibfnamefont {H.~T.}\ \bibnamefont
  {Diep}},\ }\href@noop {} {\emph {\bibinfo {title} {Frustrated spin
  systems}}}\ (\bibinfo  {publisher} {World Scientific},\ \bibinfo {year}
  {2013})\BibitemShut {NoStop}%
\bibitem [{\citenamefont {Fennell}\ \emph {et~al.}(2014)\citenamefont
  {Fennell}, \citenamefont {Kenzelmann}, \citenamefont {Roessli}, \citenamefont
  {Mutka}, \citenamefont {Ollivier}, \citenamefont {Ruminy}, \citenamefont
  {Stuhr}, \citenamefont {Zaharko}, \citenamefont {Bovo}, \citenamefont
  {Cervellino}, \citenamefont {Haas},\ and\ \citenamefont
  {Cava}}]{Fennell2014magnetoelastic}%
  \BibitemOpen
  \bibfield  {author} {\bibinfo {author} {\bibfnamefont {T.}~\bibnamefont
  {Fennell}}, \bibinfo {author} {\bibfnamefont {M.}~\bibnamefont {Kenzelmann}},
  \bibinfo {author} {\bibfnamefont {B.}~\bibnamefont {Roessli}}, \bibinfo
  {author} {\bibfnamefont {H.}~\bibnamefont {Mutka}}, \bibinfo {author}
  {\bibfnamefont {J.}~\bibnamefont {Ollivier}}, \bibinfo {author}
  {\bibfnamefont {M.}~\bibnamefont {Ruminy}}, \bibinfo {author} {\bibfnamefont
  {U.}~\bibnamefont {Stuhr}}, \bibinfo {author} {\bibfnamefont
  {O.}~\bibnamefont {Zaharko}}, \bibinfo {author} {\bibfnamefont
  {L.}~\bibnamefont {Bovo}}, \bibinfo {author} {\bibfnamefont {A.}~\bibnamefont
  {Cervellino}}, \bibinfo {author} {\bibfnamefont {M.~K.}\ \bibnamefont
  {Haas}}, \ and\ \bibinfo {author} {\bibfnamefont {R.~J.}\ \bibnamefont
  {Cava}},\ }\href {\doibase 10.1103/PhysRevLett.112.017203} {\bibfield
  {journal} {\bibinfo  {journal} {Phys. Rev. Lett.}\ }\textbf {\bibinfo
  {volume} {112}},\ \bibinfo {pages} {017203} (\bibinfo {year}
  {2014})}\BibitemShut {NoStop}%
\bibitem [{\citenamefont {Castelnovo}\ \emph {et~al.}(2008)\citenamefont
  {Castelnovo}, \citenamefont {Moessner},\ and\ \citenamefont
  {Sondhi}}]{Castelnovo}%
  \BibitemOpen
  \bibfield  {author} {\bibinfo {author} {\bibfnamefont {C.}~\bibnamefont
  {Castelnovo}}, \bibinfo {author} {\bibfnamefont {R.}~\bibnamefont
  {Moessner}}, \ and\ \bibinfo {author} {\bibfnamefont {S.~L.}\ \bibnamefont
  {Sondhi}},\ }\href {\doibase 10.1038/nature06433} {\bibfield  {journal}
  {\bibinfo  {journal} {Nature}\ }\textbf {\bibinfo {volume} {451}},\ \bibinfo
  {pages} {42} (\bibinfo {year} {2008})}\BibitemShut {NoStop}%
\bibitem [{\citenamefont {Slobinsky}\ \emph {et~al.}(2018)\citenamefont
  {Slobinsky}, \citenamefont {Baglietto},\ and\ \citenamefont
  {Borzi}}]{Slobinsky2018charge}%
  \BibitemOpen
  \bibfield  {author} {\bibinfo {author} {\bibfnamefont {D.}~\bibnamefont
  {Slobinsky}}, \bibinfo {author} {\bibfnamefont {G.}~\bibnamefont
  {Baglietto}}, \ and\ \bibinfo {author} {\bibfnamefont {R.~A.}\ \bibnamefont
  {Borzi}},\ }\href {\doibase 10.1103/PhysRevB.97.174422} {\bibfield  {journal}
  {\bibinfo  {journal} {Phys. Rev. B}\ }\textbf {\bibinfo {volume} {97}},\
  \bibinfo {pages} {174422} (\bibinfo {year} {2018})}\BibitemShut {NoStop}%
\bibitem [{\citenamefont {Borzi}\ \emph {et~al.}(2013)\citenamefont {Borzi},
  \citenamefont {Slobinsky},\ and\ \citenamefont {Grigera}}]{Borzi13}%
  \BibitemOpen
  \bibfield  {author} {\bibinfo {author} {\bibfnamefont {R.~A.}\ \bibnamefont
  {Borzi}}, \bibinfo {author} {\bibfnamefont {D.}~\bibnamefont {Slobinsky}}, \
  and\ \bibinfo {author} {\bibfnamefont {S.~A.}\ \bibnamefont {Grigera}},\
  }\href {\doibase 10.1103/PhysRevLett.111.147204} {\bibfield  {journal}
  {\bibinfo  {journal} {Phys. Rev. Lett.}\ }\textbf {\bibinfo {volume} {111}},\
  \bibinfo {pages} {147204} (\bibinfo {year} {2013})}\BibitemShut {NoStop}%
\bibitem [{\citenamefont {Guruciaga}\ \emph {et~al.}(2014)\citenamefont
  {Guruciaga}, \citenamefont {Grigera},\ and\ \citenamefont
  {Borzi}}]{Guruciaga14}%
  \BibitemOpen
  \bibfield  {author} {\bibinfo {author} {\bibfnamefont {P.~C.}\ \bibnamefont
  {Guruciaga}}, \bibinfo {author} {\bibfnamefont {S.~A.}\ \bibnamefont
  {Grigera}}, \ and\ \bibinfo {author} {\bibfnamefont {R.~A.}\ \bibnamefont
  {Borzi}},\ }\href {\doibase 10.1103/PhysRevB.90.184423} {\bibfield  {journal}
  {\bibinfo  {journal} {Phys. Rev. B}\ }\textbf {\bibinfo {volume} {90}},\
  \bibinfo {pages} {184423} (\bibinfo {year} {2014})}\BibitemShut {NoStop}%
\bibitem [{\citenamefont {Brooks-Bartlett}\ \emph {et~al.}(2014)\citenamefont
  {Brooks-Bartlett}, \citenamefont {Banks}, \citenamefont {Jaubert},
  \citenamefont {Harman-Clarke},\ and\ \citenamefont {Holdsworth}}]{Brooks14}%
  \BibitemOpen
  \bibfield  {author} {\bibinfo {author} {\bibfnamefont {M.~E.}\ \bibnamefont
  {Brooks-Bartlett}}, \bibinfo {author} {\bibfnamefont {S.~T.}\ \bibnamefont
  {Banks}}, \bibinfo {author} {\bibfnamefont {L.~D.~C.}\ \bibnamefont
  {Jaubert}}, \bibinfo {author} {\bibfnamefont {A.}~\bibnamefont
  {Harman-Clarke}}, \ and\ \bibinfo {author} {\bibfnamefont {P.~C.~W.}\
  \bibnamefont {Holdsworth}},\ }\href {\doibase 10.1103/PhysRevX.4.011007}
  {\bibfield  {journal} {\bibinfo  {journal} {Phys. Rev. X}\ }\textbf {\bibinfo
  {volume} {4}},\ \bibinfo {pages} {011007} (\bibinfo {year}
  {2014})}\BibitemShut {NoStop}%
\bibitem [{\citenamefont {Jaubert}(2015)}]{jaubert2015holes}%
  \BibitemOpen
  \bibfield  {author} {\bibinfo {author} {\bibfnamefont {L.~D.~C.}\
  \bibnamefont {Jaubert}},\ }\href {\doibase 10.1142/S2010324715400056}
  {\bibfield  {journal} {\bibinfo  {journal} {SPIN}\ }\textbf {\bibinfo
  {volume} {05}},\ \bibinfo {pages} {1540005} (\bibinfo {year}
  {2015})}\BibitemShut {NoStop}%
\bibitem [{\citenamefont {Sazonov}\ \emph {et~al.}(2012)\citenamefont
  {Sazonov}, \citenamefont {Gukasov}, \citenamefont {Mirebeau},\ and\
  \citenamefont {Bonville}}]{Sazonov2012double}%
  \BibitemOpen
  \bibfield  {author} {\bibinfo {author} {\bibfnamefont {A.~P.}\ \bibnamefont
  {Sazonov}}, \bibinfo {author} {\bibfnamefont {A.}~\bibnamefont {Gukasov}},
  \bibinfo {author} {\bibfnamefont {I.}~\bibnamefont {Mirebeau}}, \ and\
  \bibinfo {author} {\bibfnamefont {P.}~\bibnamefont {Bonville}},\ }\href
  {\doibase 10.1103/PhysRevB.85.214420} {\bibfield  {journal} {\bibinfo
  {journal} {Phys. Rev. B}\ }\textbf {\bibinfo {volume} {85}},\ \bibinfo
  {pages} {214420} (\bibinfo {year} {2012})}\BibitemShut {NoStop}%
\bibitem [{\citenamefont {Jaubert}\ and\ \citenamefont
  {Moessner}(2015)}]{Jaubert2015crystallog}%
  \BibitemOpen
  \bibfield  {author} {\bibinfo {author} {\bibfnamefont {L.~D.~C.}\
  \bibnamefont {Jaubert}}\ and\ \bibinfo {author} {\bibfnamefont
  {R.}~\bibnamefont {Moessner}},\ }\href {\doibase 10.1103/PhysRevB.91.214422}
  {\bibfield  {journal} {\bibinfo  {journal} {Phys. Rev. B}\ }\textbf {\bibinfo
  {volume} {91}},\ \bibinfo {pages} {214422} (\bibinfo {year}
  {2015})}\BibitemShut {NoStop}%
\bibitem [{\citenamefont {Borzi}\ \emph {et~al.}(2016)\citenamefont {Borzi},
  \citenamefont {G{\'o}mez~Albarrac{\'\i}n}, \citenamefont {Rosales},
  \citenamefont {Rossini}, \citenamefont {Steppke}, \citenamefont
  {Prabhakaran}, \citenamefont {Mackenzie}, \citenamefont {Cabra},\ and\
  \citenamefont {Grigera}}]{Borzi2016intermediate}%
  \BibitemOpen
  \bibfield  {author} {\bibinfo {author} {\bibfnamefont {R.~A.}\ \bibnamefont
  {Borzi}}, \bibinfo {author} {\bibfnamefont {F.~A.}\ \bibnamefont
  {G{\'o}mez~Albarrac{\'\i}n}}, \bibinfo {author} {\bibfnamefont {H.~D.}\
  \bibnamefont {Rosales}}, \bibinfo {author} {\bibfnamefont {G.~L.}\
  \bibnamefont {Rossini}}, \bibinfo {author} {\bibfnamefont {A.}~\bibnamefont
  {Steppke}}, \bibinfo {author} {\bibfnamefont {D.}~\bibnamefont
  {Prabhakaran}}, \bibinfo {author} {\bibfnamefont {A.~P.}\ \bibnamefont
  {Mackenzie}}, \bibinfo {author} {\bibfnamefont {D.~C.}\ \bibnamefont
  {Cabra}}, \ and\ \bibinfo {author} {\bibfnamefont {S.~A.}\ \bibnamefont
  {Grigera}},\ }\href {https://www.nature.com/articles/ncomms12592} {\bibfield
  {journal} {\bibinfo  {journal} {Nat. Commun.}\ }\textbf {\bibinfo {volume}
  {7}},\ \bibinfo {pages} {12592} (\bibinfo {year} {2016})}\BibitemShut
  {NoStop}%
\bibitem [{\citenamefont {Udagawa}\ \emph {et~al.}(2016)\citenamefont
  {Udagawa}, \citenamefont {Jaubert}, \citenamefont {Castelnovo},\ and\
  \citenamefont {Moessner}}]{Udagawa2016}%
  \BibitemOpen
  \bibfield  {author} {\bibinfo {author} {\bibfnamefont {M.}~\bibnamefont
  {Udagawa}}, \bibinfo {author} {\bibfnamefont {L.~D.~C.}\ \bibnamefont
  {Jaubert}}, \bibinfo {author} {\bibfnamefont {C.}~\bibnamefont {Castelnovo}},
  \ and\ \bibinfo {author} {\bibfnamefont {R.}~\bibnamefont {Moessner}},\
  }\href {\doibase 10.1103/PhysRevB.94.104416} {\bibfield  {journal} {\bibinfo
  {journal} {Phys. Rev. B}\ }\textbf {\bibinfo {volume} {94}},\ \bibinfo
  {pages} {104416} (\bibinfo {year} {2016})}\BibitemShut {NoStop}%
\bibitem [{\citenamefont {Rau}\ and\ \citenamefont
  {Gingras}(2016)}]{Rau2016spinslush}%
  \BibitemOpen
  \bibfield  {author} {\bibinfo {author} {\bibfnamefont {J.~G.}\ \bibnamefont
  {Rau}}\ and\ \bibinfo {author} {\bibfnamefont {M.~J.}\ \bibnamefont
  {Gingras}},\ }\href {https://www.nature.com/articles/ncomms12234} {\bibfield
  {journal} {\bibinfo  {journal} {Nat. Commun.}\ }\textbf {\bibinfo {volume}
  {7}},\ \bibinfo {pages} {12234} (\bibinfo {year} {2016})}\BibitemShut
  {NoStop}%
\bibitem [{\citenamefont {Xie}\ \emph {et~al.}(2015)\citenamefont {Xie},
  \citenamefont {Du}, \citenamefont {Yan},\ and\ \citenamefont
  {Liu}}]{Xie2015magnetic}%
  \BibitemOpen
  \bibfield  {author} {\bibinfo {author} {\bibfnamefont {Y.-L.}\ \bibnamefont
  {Xie}}, \bibinfo {author} {\bibfnamefont {Z.-Z.}\ \bibnamefont {Du}},
  \bibinfo {author} {\bibfnamefont {Z.-B.}\ \bibnamefont {Yan}}, \ and\
  \bibinfo {author} {\bibfnamefont {J.-M.}\ \bibnamefont {Liu}},\ }\href
  {https://www.nature.com/articles/srep15875} {\bibfield  {journal} {\bibinfo
  {journal} {Sci. Rep.}\ }\textbf {\bibinfo {volume} {5}},\ \bibinfo {pages}
  {15875} (\bibinfo {year} {2015})}\BibitemShut {NoStop}%
\bibitem [{\citenamefont {Farhan}\ \emph {et~al.}(2019)\citenamefont {Farhan},
  \citenamefont {Saccone}, \citenamefont {Petersen}, \citenamefont {Dhuey},
  \citenamefont {Chopdekar}, \citenamefont {Huang}, \citenamefont {Kent},
  \citenamefont {Chen}, \citenamefont {Alava}, \citenamefont {Lippert},
  \citenamefont {Scholl},\ and\ \citenamefont {van Dijken}}]{Farhaneaav6380}%
  \BibitemOpen
  \bibfield  {author} {\bibinfo {author} {\bibfnamefont {A.}~\bibnamefont
  {Farhan}}, \bibinfo {author} {\bibfnamefont {M.}~\bibnamefont {Saccone}},
  \bibinfo {author} {\bibfnamefont {C.~F.}\ \bibnamefont {Petersen}}, \bibinfo
  {author} {\bibfnamefont {S.}~\bibnamefont {Dhuey}}, \bibinfo {author}
  {\bibfnamefont {R.~V.}\ \bibnamefont {Chopdekar}}, \bibinfo {author}
  {\bibfnamefont {Y.-L.}\ \bibnamefont {Huang}}, \bibinfo {author}
  {\bibfnamefont {N.}~\bibnamefont {Kent}}, \bibinfo {author} {\bibfnamefont
  {Z.}~\bibnamefont {Chen}}, \bibinfo {author} {\bibfnamefont {M.~J.}\
  \bibnamefont {Alava}}, \bibinfo {author} {\bibfnamefont {T.}~\bibnamefont
  {Lippert}}, \bibinfo {author} {\bibfnamefont {A.}~\bibnamefont {Scholl}}, \
  and\ \bibinfo {author} {\bibfnamefont {S.}~\bibnamefont {van Dijken}},\
  }\href {http://advances.sciencemag.org/content/5/2/eaav6380} {\bibfield
  {journal} {\bibinfo  {journal} {Sci. Adv.}\ }\textbf {\bibinfo {volume} {5}}
  (\bibinfo {year} {2019})}\BibitemShut {NoStop}%
\bibitem [{\citenamefont {Reichhardt}\ \emph {et~al.}(2012)\citenamefont
  {Reichhardt}, \citenamefont {Lib{\'{a}}l},\ and\ \citenamefont
  {Reichhardt}}]{Olson_Reichhardt_2012}%
  \BibitemOpen
  \bibfield  {author} {\bibinfo {author} {\bibfnamefont {C.~J.~O.}\
  \bibnamefont {Reichhardt}}, \bibinfo {author} {\bibfnamefont
  {A.}~\bibnamefont {Lib{\'{a}}l}}, \ and\ \bibinfo {author} {\bibfnamefont
  {C.}~\bibnamefont {Reichhardt}},\ }\href {\doibase
  10.1088/1367-2630/14/2/025006} {\bibfield  {journal} {\bibinfo  {journal}
  {New J. Phys.}\ }\textbf {\bibinfo {volume} {14}},\ \bibinfo {pages} {025006}
  (\bibinfo {year} {2012})}\BibitemShut {NoStop}%
\bibitem [{\citenamefont {Chalker}(2011)}]{Chalker}%
  \BibitemOpen
  \bibfield  {author} {\bibinfo {author} {\bibfnamefont {J.~T.}\ \bibnamefont
  {Chalker}},\ }in\ \href@noop {} {\emph {\bibinfo {booktitle} {Introduction to
  Frustrated Magnetism: Materials, Experiments, Theory}}},\ \bibinfo {editor}
  {edited by\ \bibinfo {editor} {\bibfnamefont {C.}~\bibnamefont {Lacroix}},
  \bibinfo {editor} {\bibfnamefont {P.}~\bibnamefont {Mendels}}, \ and\
  \bibinfo {editor} {\bibfnamefont {F.}~\bibnamefont {Mila}}}\ (\bibinfo
  {publisher} {Springer Science \& Business Media},\ \bibinfo {year} {2011})\
  Chap.\ \bibinfo {chapter} {Geometrically frustrated antiferromagnets:
  statistical mechanics and dynamics}\BibitemShut {NoStop}%
\bibitem [{\citenamefont {Moessner}\ and\ \citenamefont
  {Chalker}(1998)}]{moessner1998low}%
  \BibitemOpen
  \bibfield  {author} {\bibinfo {author} {\bibfnamefont {R.}~\bibnamefont
  {Moessner}}\ and\ \bibinfo {author} {\bibfnamefont {J.~T.}\ \bibnamefont
  {Chalker}},\ }\href {\doibase 10.1103/PhysRevB.58.12049} {\bibfield
  {journal} {\bibinfo  {journal} {Phys. Rev. B}\ }\textbf {\bibinfo {volume}
  {58}},\ \bibinfo {pages} {12049} (\bibinfo {year} {1998})}\BibitemShut
  {NoStop}%
\bibitem [{\citenamefont {Shender}\ and\ \citenamefont
  {Holdsworth}(1996)}]{Shender1996review}%
  \BibitemOpen
  \bibfield  {author} {\bibinfo {author} {\bibfnamefont {E.~F.}\ \bibnamefont
  {Shender}}\ and\ \bibinfo {author} {\bibfnamefont {P.~C.~W.}\ \bibnamefont
  {Holdsworth}},\ }in\ \href@noop {} {\emph {\bibinfo {booktitle} {Fluctuations
  and Order}}}\ (\bibinfo  {publisher} {Springer},\ \bibinfo {year} {1996})\
  Chap.\ \bibinfo {chapter} {Order by disorder and topology in frustrated
  magnetic systems}\BibitemShut {NoStop}%
\bibitem [{\citenamefont {Bramwell}\ \emph {et~al.}(1994)\citenamefont
  {Bramwell}, \citenamefont {Gingras},\ and\ \citenamefont
  {Reimers}}]{bramwell1994order}%
  \BibitemOpen
  \bibfield  {author} {\bibinfo {author} {\bibfnamefont {S.~T.}\ \bibnamefont
  {Bramwell}}, \bibinfo {author} {\bibfnamefont {M.~J.~P.}\ \bibnamefont
  {Gingras}}, \ and\ \bibinfo {author} {\bibfnamefont {J.~N.}\ \bibnamefont
  {Reimers}},\ }\href {\doibase 10.1063/1.355676} {\bibfield  {journal}
  {\bibinfo  {journal} {J. Appl. Phys.}\ }\textbf {\bibinfo {volume} {75}},\
  \bibinfo {pages} {5523} (\bibinfo {year} {1994})}\BibitemShut {NoStop}%
\bibitem [{\citenamefont {Villain}\ \emph {et~al.}(1980)\citenamefont
  {Villain}, \citenamefont {Bidaux}, \citenamefont {Carton},\ and\
  \citenamefont {Conte}}]{Villain}%
  \BibitemOpen
  \bibfield  {author} {\bibinfo {author} {\bibfnamefont {J.}~\bibnamefont
  {Villain}}, \bibinfo {author} {\bibfnamefont {R.}~\bibnamefont {Bidaux}},
  \bibinfo {author} {\bibfnamefont {J.-P.}\ \bibnamefont {Carton}}, \ and\
  \bibinfo {author} {\bibfnamefont {R.}~\bibnamefont {Conte}},\ }\href
  {\doibase 10.1051/jphys:0198000410110126300} {\bibfield  {journal} {\bibinfo
  {journal} {J. Phys.}\ }\textbf {\bibinfo {volume} {41}},\ \bibinfo {pages}
  {1263} (\bibinfo {year} {1980})}\BibitemShut {NoStop}%
\bibitem [{\citenamefont {Maryasin}\ and\ \citenamefont
  {Zhitomirsky}(2014)}]{Maryasin2014structural}%
  \BibitemOpen
  \bibfield  {author} {\bibinfo {author} {\bibfnamefont {V.~S.}\ \bibnamefont
  {Maryasin}}\ and\ \bibinfo {author} {\bibfnamefont {M.~E.}\ \bibnamefont
  {Zhitomirsky}},\ }\href {\doibase 10.1103/PhysRevB.90.094412} {\bibfield
  {journal} {\bibinfo  {journal} {Phys. Rev. B}\ }\textbf {\bibinfo {volume}
  {90}},\ \bibinfo {pages} {094412} (\bibinfo {year} {2014})}\BibitemShut
  {NoStop}%
\bibitem [{\citenamefont {Henley}(1989)}]{Henley1989ordering}%
  \BibitemOpen
  \bibfield  {author} {\bibinfo {author} {\bibfnamefont {C.~L.}\ \bibnamefont
  {Henley}},\ }\href {\doibase 10.1103/PhysRevLett.62.2056} {\bibfield
  {journal} {\bibinfo  {journal} {Phys. Rev. Lett.}\ }\textbf {\bibinfo
  {volume} {62}},\ \bibinfo {pages} {2056} (\bibinfo {year}
  {1989})}\BibitemShut {NoStop}%
\bibitem [{\citenamefont {Savary}\ \emph {et~al.}(2012)\citenamefont {Savary},
  \citenamefont {Ross}, \citenamefont {Gaulin}, \citenamefont {Ruff},\ and\
  \citenamefont {Balents}}]{Savary12}%
  \BibitemOpen
  \bibfield  {author} {\bibinfo {author} {\bibfnamefont {L.}~\bibnamefont
  {Savary}}, \bibinfo {author} {\bibfnamefont {K.~A.}\ \bibnamefont {Ross}},
  \bibinfo {author} {\bibfnamefont {B.~D.}\ \bibnamefont {Gaulin}}, \bibinfo
  {author} {\bibfnamefont {J.~P.~C.}\ \bibnamefont {Ruff}}, \ and\ \bibinfo
  {author} {\bibfnamefont {L.}~\bibnamefont {Balents}},\ }\href {\doibase
  10.1103/PhysRevLett.109.167201} {\bibfield  {journal} {\bibinfo  {journal}
  {Phys. Rev. Lett.}\ }\textbf {\bibinfo {volume} {109}},\ \bibinfo {pages}
  {167201} (\bibinfo {year} {2012})}\BibitemShut {NoStop}%
\bibitem [{\citenamefont {Rau}\ \emph {et~al.}(2016)\citenamefont {Rau},
  \citenamefont {Petit},\ and\ \citenamefont {Gingras}}]{Rau2016order}%
  \BibitemOpen
  \bibfield  {author} {\bibinfo {author} {\bibfnamefont {J.~G.}\ \bibnamefont
  {Rau}}, \bibinfo {author} {\bibfnamefont {S.}~\bibnamefont {Petit}}, \ and\
  \bibinfo {author} {\bibfnamefont {M.~J.~P.}\ \bibnamefont {Gingras}},\ }\href
  {\doibase 10.1103/PhysRevB.93.184408} {\bibfield  {journal} {\bibinfo
  {journal} {Phys. Rev. B}\ }\textbf {\bibinfo {volume} {93}},\ \bibinfo
  {pages} {184408} (\bibinfo {year} {2016})}\BibitemShut {NoStop}%
\bibitem [{\citenamefont {Zhitomirsky}\ \emph {et~al.}(2012)\citenamefont
  {Zhitomirsky}, \citenamefont {Gvozdikova}, \citenamefont {Holdsworth},\ and\
  \citenamefont {Moessner}}]{Zhitomirsky12}%
  \BibitemOpen
  \bibfield  {author} {\bibinfo {author} {\bibfnamefont {M.~E.}\ \bibnamefont
  {Zhitomirsky}}, \bibinfo {author} {\bibfnamefont {M.~V.}\ \bibnamefont
  {Gvozdikova}}, \bibinfo {author} {\bibfnamefont {P.~C.~W.}\ \bibnamefont
  {Holdsworth}}, \ and\ \bibinfo {author} {\bibfnamefont {R.}~\bibnamefont
  {Moessner}},\ }\href
  {http://journals.aps.org/prl/abstract/10.1103/PhysRevLett.109.077204}
  {\bibfield  {journal} {\bibinfo  {journal} {Phys. Rev. Lett.}\ }\textbf
  {\bibinfo {volume} {109}},\ \bibinfo {pages} {077204} (\bibinfo {year}
  {2012})}\BibitemShut {NoStop}%
\bibitem [{\citenamefont {Ross}\ \emph {et~al.}(2014)\citenamefont {Ross},
  \citenamefont {Qiu}, \citenamefont {Copley}, \citenamefont {Dabkowska},\ and\
  \citenamefont {Gaulin}}]{Ross14}%
  \BibitemOpen
  \bibfield  {author} {\bibinfo {author} {\bibfnamefont {K.~A.}\ \bibnamefont
  {Ross}}, \bibinfo {author} {\bibfnamefont {Y.}~\bibnamefont {Qiu}}, \bibinfo
  {author} {\bibfnamefont {J.~R.~D.}\ \bibnamefont {Copley}}, \bibinfo {author}
  {\bibfnamefont {H.~A.}\ \bibnamefont {Dabkowska}}, \ and\ \bibinfo {author}
  {\bibfnamefont {B.~D.}\ \bibnamefont {Gaulin}},\ }\href
  {http://journals.aps.org/prl/abstract/10.1103/PhysRevLett.112.057201}
  {\bibfield  {journal} {\bibinfo  {journal} {Phys. Rev. Lett.}\ }\textbf
  {\bibinfo {volume} {112}},\ \bibinfo {pages} {057201} (\bibinfo {year}
  {2014})}\BibitemShut {NoStop}%
\bibitem [{\citenamefont {Champion}\ \emph {et~al.}(2003)\citenamefont
  {Champion}, \citenamefont {Harris}, \citenamefont {Holdsworth}, \citenamefont
  {Wills}, \citenamefont {Balakrishnan}, \citenamefont {Bramwell},
  \citenamefont {Cizmar}, \citenamefont {Fennell}, \citenamefont {Gardner},
  \citenamefont {Lago}, \citenamefont {McMorrow}, \citenamefont {Orendac},
  \citenamefont {Orendacova}, \citenamefont {McK.~Paul}, \citenamefont {Smith},
  \citenamefont {Telling},\ and\ \citenamefont {Wildes}}]{Champion03}%
  \BibitemOpen
  \bibfield  {author} {\bibinfo {author} {\bibfnamefont {J.~D.~M.}\
  \bibnamefont {Champion}}, \bibinfo {author} {\bibfnamefont {M.~J.}\
  \bibnamefont {Harris}}, \bibinfo {author} {\bibfnamefont {P.~C.~W.}\
  \bibnamefont {Holdsworth}}, \bibinfo {author} {\bibfnamefont {A.~S.}\
  \bibnamefont {Wills}}, \bibinfo {author} {\bibfnamefont {G.}~\bibnamefont
  {Balakrishnan}}, \bibinfo {author} {\bibfnamefont {S.~T.}\ \bibnamefont
  {Bramwell}}, \bibinfo {author} {\bibfnamefont {E.}~\bibnamefont {Cizmar}},
  \bibinfo {author} {\bibfnamefont {T.}~\bibnamefont {Fennell}}, \bibinfo
  {author} {\bibfnamefont {J.~S.}\ \bibnamefont {Gardner}}, \bibinfo {author}
  {\bibfnamefont {J.}~\bibnamefont {Lago}}, \bibinfo {author} {\bibfnamefont
  {D.~F.}\ \bibnamefont {McMorrow}}, \bibinfo {author} {\bibfnamefont
  {M.}~\bibnamefont {Orendac}}, \bibinfo {author} {\bibfnamefont
  {A.}~\bibnamefont {Orendacova}}, \bibinfo {author} {\bibfnamefont
  {D.}~\bibnamefont {McK.~Paul}}, \bibinfo {author} {\bibfnamefont {R.~I.}\
  \bibnamefont {Smith}}, \bibinfo {author} {\bibfnamefont {M.~T.~F.}\
  \bibnamefont {Telling}}, \ and\ \bibinfo {author} {\bibfnamefont
  {A.}~\bibnamefont {Wildes}},\ }\href {\doibase 10.1103/PhysRevB.68.020401}
  {\bibfield  {journal} {\bibinfo  {journal} {Phys. Rev. B}\ }\textbf {\bibinfo
  {volume} {68}},\ \bibinfo {pages} {020401(R)} (\bibinfo {year}
  {2003})}\BibitemShut {NoStop}%
\bibitem [{\citenamefont {McClarty}\ \emph {et~al.}(2014)\citenamefont
  {McClarty}, \citenamefont {Stasiak},\ and\ \citenamefont
  {Gingras}}]{Mcclarty2014order}%
  \BibitemOpen
  \bibfield  {author} {\bibinfo {author} {\bibfnamefont {P.~A.}\ \bibnamefont
  {McClarty}}, \bibinfo {author} {\bibfnamefont {P.}~\bibnamefont {Stasiak}}, \
  and\ \bibinfo {author} {\bibfnamefont {M.~J.~P.}\ \bibnamefont {Gingras}},\
  }\href {\doibase 10.1103/PhysRevB.89.024425} {\bibfield  {journal} {\bibinfo
  {journal} {Phys. Rev. B}\ }\textbf {\bibinfo {volume} {89}},\ \bibinfo
  {pages} {024425} (\bibinfo {year} {2014})}\BibitemShut {NoStop}%
\bibitem [{\citenamefont {Oitmaa}\ \emph {et~al.}(2013)\citenamefont {Oitmaa},
  \citenamefont {Singh}, \citenamefont {Javanparast}, \citenamefont {Day},
  \citenamefont {Bagheri},\ and\ \citenamefont {Gingras}}]{Oitmaa13}%
  \BibitemOpen
  \bibfield  {author} {\bibinfo {author} {\bibfnamefont {J.}~\bibnamefont
  {Oitmaa}}, \bibinfo {author} {\bibfnamefont {R.~R.~P.}\ \bibnamefont
  {Singh}}, \bibinfo {author} {\bibfnamefont {B.}~\bibnamefont {Javanparast}},
  \bibinfo {author} {\bibfnamefont {A.~G.~R.}\ \bibnamefont {Day}}, \bibinfo
  {author} {\bibfnamefont {B.~V.}\ \bibnamefont {Bagheri}}, \ and\ \bibinfo
  {author} {\bibfnamefont {M.~J.~P.}\ \bibnamefont {Gingras}},\ }\href
  {\doibase 10.1103/PhysRevB.88.220404} {\bibfield  {journal} {\bibinfo
  {journal} {Phys. Rev. B}\ }\textbf {\bibinfo {volume} {88}},\ \bibinfo
  {pages} {220404(R)} (\bibinfo {year} {2013})}\BibitemShut {NoStop}%
\bibitem [{\citenamefont {Guruciaga}\ \emph {et~al.}(2016)\citenamefont
  {Guruciaga}, \citenamefont {Tarzia}, \citenamefont {Ferreyra}, \citenamefont
  {Cugliandolo}, \citenamefont {Grigera},\ and\ \citenamefont
  {Borzi}}]{Guruciaga16}%
  \BibitemOpen
  \bibfield  {author} {\bibinfo {author} {\bibfnamefont {P.~C.}\ \bibnamefont
  {Guruciaga}}, \bibinfo {author} {\bibfnamefont {M.}~\bibnamefont {Tarzia}},
  \bibinfo {author} {\bibfnamefont {M.~V.}\ \bibnamefont {Ferreyra}}, \bibinfo
  {author} {\bibfnamefont {L.~F.}\ \bibnamefont {Cugliandolo}}, \bibinfo
  {author} {\bibfnamefont {S.~A.}\ \bibnamefont {Grigera}}, \ and\ \bibinfo
  {author} {\bibfnamefont {R.~A.}\ \bibnamefont {Borzi}},\ }\href {\doibase
  10.1103/PhysRevLett.117.167203} {\bibfield  {journal} {\bibinfo  {journal}
  {Phys. Rev. Lett.}\ }\textbf {\bibinfo {volume} {117}},\ \bibinfo {pages}
  {167203} (\bibinfo {year} {2016})}\BibitemShut {NoStop}%
\bibitem [{\citenamefont {Sleight}\ \emph {et~al.}(1974)\citenamefont
  {Sleight}, \citenamefont {Gillson}, \citenamefont {Weiher},\ and\
  \citenamefont {Bindloss}}]{Sleight1974semiconductor}%
  \BibitemOpen
  \bibfield  {author} {\bibinfo {author} {\bibfnamefont {A.~W.}\ \bibnamefont
  {Sleight}}, \bibinfo {author} {\bibfnamefont {J.~L.}\ \bibnamefont
  {Gillson}}, \bibinfo {author} {\bibfnamefont {J.~F.}\ \bibnamefont {Weiher}},
  \ and\ \bibinfo {author} {\bibfnamefont {W.}~\bibnamefont {Bindloss}},\
  }\href {https://www.sciencedirect.com/science/article/pii/003810987490917X}
  {\bibfield  {journal} {\bibinfo  {journal} {Solid State Commun.}\ }\textbf
  {\bibinfo {volume} {14}},\ \bibinfo {pages} {357} (\bibinfo {year}
  {1974})}\BibitemShut {NoStop}%
\bibitem [{\citenamefont {Lhotel}\ \emph {et~al.}(2015)\citenamefont {Lhotel},
  \citenamefont {Petit}, \citenamefont {Guitteny}, \citenamefont {Florea},
  \citenamefont {Ciomaga~Hatnean}, \citenamefont {Colin}, \citenamefont
  {Ressouche}, \citenamefont {Lees},\ and\ \citenamefont
  {Balakrishnan}}]{Lhotel2015fluctuations}%
  \BibitemOpen
  \bibfield  {author} {\bibinfo {author} {\bibfnamefont {E.}~\bibnamefont
  {Lhotel}}, \bibinfo {author} {\bibfnamefont {S.}~\bibnamefont {Petit}},
  \bibinfo {author} {\bibfnamefont {S.}~\bibnamefont {Guitteny}}, \bibinfo
  {author} {\bibfnamefont {O.}~\bibnamefont {Florea}}, \bibinfo {author}
  {\bibfnamefont {M.}~\bibnamefont {Ciomaga~Hatnean}}, \bibinfo {author}
  {\bibfnamefont {C.}~\bibnamefont {Colin}}, \bibinfo {author} {\bibfnamefont
  {E.}~\bibnamefont {Ressouche}}, \bibinfo {author} {\bibfnamefont {M.~R.}\
  \bibnamefont {Lees}}, \ and\ \bibinfo {author} {\bibfnamefont
  {G.}~\bibnamefont {Balakrishnan}},\ }\href {\doibase
  10.1103/PhysRevLett.115.197202} {\bibfield  {journal} {\bibinfo  {journal}
  {Phys. Rev. Lett.}\ }\textbf {\bibinfo {volume} {115}},\ \bibinfo {pages}
  {197202} (\bibinfo {year} {2015})}\BibitemShut {NoStop}%
\bibitem [{\citenamefont {Xu}\ \emph {et~al.}(2015)\citenamefont {Xu},
  \citenamefont {Anand}, \citenamefont {Bera}, \citenamefont {Frontzek},
  \citenamefont {Abernathy}, \citenamefont {Casati}, \citenamefont
  {Siemensmeyer},\ and\ \citenamefont {Lake}}]{Xu2015}%
  \BibitemOpen
  \bibfield  {author} {\bibinfo {author} {\bibfnamefont {J.}~\bibnamefont
  {Xu}}, \bibinfo {author} {\bibfnamefont {V.~K.}\ \bibnamefont {Anand}},
  \bibinfo {author} {\bibfnamefont {A.~K.}\ \bibnamefont {Bera}}, \bibinfo
  {author} {\bibfnamefont {M.}~\bibnamefont {Frontzek}}, \bibinfo {author}
  {\bibfnamefont {D.~L.}\ \bibnamefont {Abernathy}}, \bibinfo {author}
  {\bibfnamefont {N.}~\bibnamefont {Casati}}, \bibinfo {author} {\bibfnamefont
  {K.}~\bibnamefont {Siemensmeyer}}, \ and\ \bibinfo {author} {\bibfnamefont
  {B.}~\bibnamefont {Lake}},\ }\href
  {http://journals.aps.org/prb/abstract/10.1103/PhysRevB.92.224430} {\bibfield
  {journal} {\bibinfo  {journal} {Phys. Rev. B}\ }\textbf {\bibinfo {volume}
  {92}},\ \bibinfo {pages} {224430} (\bibinfo {year} {2015})}\BibitemShut
  {NoStop}%
\bibitem [{\citenamefont {Anand}\ \emph {et~al.}(2015)\citenamefont {Anand},
  \citenamefont {Bera}, \citenamefont {Xu}, \citenamefont {Herrmannsd\"orfer},
  \citenamefont {Ritter},\ and\ \citenamefont {Lake}}]{Anand15}%
  \BibitemOpen
  \bibfield  {author} {\bibinfo {author} {\bibfnamefont {V.~K.}\ \bibnamefont
  {Anand}}, \bibinfo {author} {\bibfnamefont {A.~K.}\ \bibnamefont {Bera}},
  \bibinfo {author} {\bibfnamefont {J.}~\bibnamefont {Xu}}, \bibinfo {author}
  {\bibfnamefont {T.}~\bibnamefont {Herrmannsd\"orfer}}, \bibinfo {author}
  {\bibfnamefont {C.}~\bibnamefont {Ritter}}, \ and\ \bibinfo {author}
  {\bibfnamefont {B.}~\bibnamefont {Lake}},\ }\href
  {http://journals.aps.org/prb/abstract/10.1103/PhysRevB.92.184418} {\bibfield
  {journal} {\bibinfo  {journal} {Phys. Rev. B}\ }\textbf {\bibinfo {volume}
  {92}},\ \bibinfo {pages} {184418} (\bibinfo {year} {2015})}\BibitemShut
  {NoStop}%
\bibitem [{\citenamefont {Bertin}\ \emph {et~al.}(2015)\citenamefont {Bertin},
  \citenamefont {Dalmas~de R\'eotier}, \citenamefont {F\aa{}k}, \citenamefont
  {Marin}, \citenamefont {Yaouanc}, \citenamefont {Forget}, \citenamefont
  {Sheptyakov}, \citenamefont {Frick}, \citenamefont {Ritter}, \citenamefont
  {Amato}, \citenamefont {Baines},\ and\ \citenamefont {King}}]{Bertin2015nd}%
  \BibitemOpen
  \bibfield  {author} {\bibinfo {author} {\bibfnamefont {A.}~\bibnamefont
  {Bertin}}, \bibinfo {author} {\bibfnamefont {P.}~\bibnamefont {Dalmas~de
  R\'eotier}}, \bibinfo {author} {\bibfnamefont {B.}~\bibnamefont {F\aa{}k}},
  \bibinfo {author} {\bibfnamefont {C.}~\bibnamefont {Marin}}, \bibinfo
  {author} {\bibfnamefont {A.}~\bibnamefont {Yaouanc}}, \bibinfo {author}
  {\bibfnamefont {A.}~\bibnamefont {Forget}}, \bibinfo {author} {\bibfnamefont
  {D.}~\bibnamefont {Sheptyakov}}, \bibinfo {author} {\bibfnamefont
  {B.}~\bibnamefont {Frick}}, \bibinfo {author} {\bibfnamefont
  {C.}~\bibnamefont {Ritter}}, \bibinfo {author} {\bibfnamefont
  {A.}~\bibnamefont {Amato}}, \bibinfo {author} {\bibfnamefont
  {C.}~\bibnamefont {Baines}}, \ and\ \bibinfo {author} {\bibfnamefont
  {P.~J.~C.}\ \bibnamefont {King}},\ }\href {\doibase
  10.1103/PhysRevB.92.144423} {\bibfield  {journal} {\bibinfo  {journal} {Phys.
  Rev. B}\ }\textbf {\bibinfo {volume} {92}},\ \bibinfo {pages} {144423}
  (\bibinfo {year} {2015})}\BibitemShut {NoStop}%
\bibitem [{\citenamefont {Hiroi}\ \emph {et~al.}(2003)\citenamefont {Hiroi},
  \citenamefont {Matsuhira},\ and\ \citenamefont {Ogata}}]{Hiroi03}%
  \BibitemOpen
  \bibfield  {author} {\bibinfo {author} {\bibfnamefont {Z.}~\bibnamefont
  {Hiroi}}, \bibinfo {author} {\bibfnamefont {K.}~\bibnamefont {Matsuhira}}, \
  and\ \bibinfo {author} {\bibfnamefont {M.}~\bibnamefont {Ogata}},\ }\href
  {http://dx.doi.org/10.1143/JPSJ.72.3045} {\bibfield  {journal} {\bibinfo
  {journal} {J. Phys. Soc. Jpn.}\ }\textbf {\bibinfo {volume} {72}},\ \bibinfo
  {pages} {3045} (\bibinfo {year} {2003})}\BibitemShut {NoStop}%
\bibitem [{\citenamefont {Melko}\ and\ \citenamefont {Gingras}(2004)}]{Melko}%
  \BibitemOpen
  \bibfield  {author} {\bibinfo {author} {\bibfnamefont {R.~G.}\ \bibnamefont
  {Melko}}\ and\ \bibinfo {author} {\bibfnamefont {M.~J.~P.}\ \bibnamefont
  {Gingras}},\ }\href {\doibase 10.1088/0953-8984/16/43/R02} {\bibfield
  {journal} {\bibinfo  {journal} {J. Phys.: Cond. Matter}\ }\textbf {\bibinfo
  {volume} {16}},\ \bibinfo {pages} {R1277} (\bibinfo {year}
  {2004})}\BibitemShut {NoStop}%
\bibitem [{Note1()}]{Note1}%
  \BibitemOpen
  \bibinfo {note} {Note that, differently from the staggered charge density
  defined below, these are \protect \emph {number} densities, irrespective of
  the monopole charges (which always average to zero).}\BibitemShut {Stop}%
\bibitem [{\citenamefont {Ashcroft}\ and\ \citenamefont
  {Mermin}(1976)}]{Ashcroft1976solid}%
  \BibitemOpen
  \bibfield  {author} {\bibinfo {author} {\bibfnamefont {N.~W.}\ \bibnamefont
  {Ashcroft}}\ and\ \bibinfo {author} {\bibfnamefont {N.~D.}\ \bibnamefont
  {Mermin}},\ }\href@noop {} {\emph {\bibinfo {title} {Solid State Physics}}}\
  (\bibinfo  {publisher} {Holt, Rinehart and Winston},\ \bibinfo {year}
  {1976})\BibitemShut {NoStop}%
\bibitem [{Note2()}]{Note2}%
  \BibitemOpen
  \bibinfo {note} {Although the lowest-possible temperatures are desirable in
  order to determine the ground-state transition field $h_{\protect \textit
  {inf}}$, compromises have to be made not only to ensure equilibrium but also
  to resolve $h_{\protect \textit {sup}}$, which involves excitations and is
  thus extremely subtle.}\BibitemShut {Stop}%
\bibitem [{Note3()}]{Note3}%
  \BibitemOpen
  \bibinfo {note} {We will see in the next section that the density of double
  monopoles could be affected in a non-trivial way by very peculiar finite-size
  effects that induce disorder at very low temperatures for $D=0$~\cite
  {Guruciaga16}. However, the system remains ordered even at $T=0$ for $D\not
  =0$, which is the important case here.}\BibitemShut {Stop}%
\bibitem [{Note4()}]{Note4}%
  \BibitemOpen
  \bibinfo {note} {The measurement of magnetocapacitance may be an interesting
  exception to this statement, {allowing an indirect approach to the formation
  of staggered monopole order} (see Refs.~\protect \rev@citealpnum
  {Katsufuji2004magnetocapacitance,Saito2005magnetodielectric,Khomskii2012electric}).}\BibitemShut
  {Stop}%
\bibitem [{\citenamefont {Anand}\ \emph {et~al.}(2017)\citenamefont {Anand},
  \citenamefont {Abernathy}, \citenamefont {Adroja}, \citenamefont {Hillier},
  \citenamefont {Biswas},\ and\ \citenamefont {Lake}}]{Anand2017muon}%
  \BibitemOpen
  \bibfield  {author} {\bibinfo {author} {\bibfnamefont {V.~K.}\ \bibnamefont
  {Anand}}, \bibinfo {author} {\bibfnamefont {D.~L.}\ \bibnamefont
  {Abernathy}}, \bibinfo {author} {\bibfnamefont {D.~T.}\ \bibnamefont
  {Adroja}}, \bibinfo {author} {\bibfnamefont {A.~D.}\ \bibnamefont {Hillier}},
  \bibinfo {author} {\bibfnamefont {P.~K.}\ \bibnamefont {Biswas}}, \ and\
  \bibinfo {author} {\bibfnamefont {B.}~\bibnamefont {Lake}},\ }\href {\doibase
  10.1103/PhysRevB.95.224420} {\bibfield  {journal} {\bibinfo  {journal} {Phys.
  Rev. B}\ }\textbf {\bibinfo {volume} {95}},\ \bibinfo {pages} {224420}
  (\bibinfo {year} {2017})}\BibitemShut {NoStop}%
\bibitem [{\citenamefont {Yamaura}\ \emph {et~al.}(2012)\citenamefont
  {Yamaura}, \citenamefont {Ohsumi}, \citenamefont {Sugimoto}, \citenamefont
  {Tsutsui}, \citenamefont {Yoda}, \citenamefont {Takeshita}, \citenamefont
  {Tokuda}, \citenamefont {Kitao}, \citenamefont {Kurokuzu}, \citenamefont
  {Seto}, \citenamefont {Yamauchi}, \citenamefont {Ohgushi}, \citenamefont
  {Takigawa}, \citenamefont {Arima},\ and\ \citenamefont
  {Hiroi}}]{Yamaura2012phase}%
  \BibitemOpen
  \bibfield  {author} {\bibinfo {author} {\bibfnamefont {J.}~\bibnamefont
  {Yamaura}}, \bibinfo {author} {\bibfnamefont {H.}~\bibnamefont {Ohsumi}},
  \bibinfo {author} {\bibfnamefont {K.}~\bibnamefont {Sugimoto}}, \bibinfo
  {author} {\bibfnamefont {S.}~\bibnamefont {Tsutsui}}, \bibinfo {author}
  {\bibfnamefont {Y.}~\bibnamefont {Yoda}}, \bibinfo {author} {\bibfnamefont
  {S.}~\bibnamefont {Takeshita}}, \bibinfo {author} {\bibfnamefont
  {A.}~\bibnamefont {Tokuda}}, \bibinfo {author} {\bibfnamefont
  {S.}~\bibnamefont {Kitao}}, \bibinfo {author} {\bibfnamefont
  {M.}~\bibnamefont {Kurokuzu}}, \bibinfo {author} {\bibfnamefont
  {M.}~\bibnamefont {Seto}}, \bibinfo {author} {\bibfnamefont {I.}~\bibnamefont
  {Yamauchi}}, \bibinfo {author} {\bibfnamefont {K.}~\bibnamefont {Ohgushi}},
  \bibinfo {author} {\bibfnamefont {M.}~\bibnamefont {Takigawa}}, \bibinfo
  {author} {\bibfnamefont {T.}~\bibnamefont {Arima}}, \ and\ \bibinfo {author}
  {\bibfnamefont {Z.}~\bibnamefont {Hiroi}},\ }\href {\doibase
  10.1088/1742-6596/391/1/012112} {\bibfield  {journal} {\bibinfo  {journal}
  {J. Phys. Conf. Ser.}\ }\textbf {\bibinfo {volume} {391}},\ \bibinfo {pages}
  {012112} (\bibinfo {year} {2012})}\BibitemShut {NoStop}%
\bibitem [{Note100()}]{Note100}%
  \BibitemOpen
  \bibinfo {note} {The $[110]$ direction has the special feature of being just
  midway between $[111]$ and $[11\protect \overline {1}]$; the latter promotes
  the opposite order.}\BibitemShut {Stop}%
\bibitem [{\citenamefont {Sakakibara}\ \emph {et~al.}(2003)\citenamefont
  {Sakakibara}, \citenamefont {Tayama}, \citenamefont {Hiroi}, \citenamefont
  {Matsuhira},\ and\ \citenamefont {Takagi}}]{Sakakibara03}%
  \BibitemOpen
  \bibfield  {author} {\bibinfo {author} {\bibfnamefont {T.}~\bibnamefont
  {Sakakibara}}, \bibinfo {author} {\bibfnamefont {T.}~\bibnamefont {Tayama}},
  \bibinfo {author} {\bibfnamefont {Z.}~\bibnamefont {Hiroi}}, \bibinfo
  {author} {\bibfnamefont {K.}~\bibnamefont {Matsuhira}}, \ and\ \bibinfo
  {author} {\bibfnamefont {S.}~\bibnamefont {Takagi}},\ }\href {\doibase
  10.1103/PhysRevLett.90.207205} {\bibfield  {journal} {\bibinfo  {journal}
  {Phys. Rev. Lett.}\ }\textbf {\bibinfo {volume} {90}},\ \bibinfo {pages}
  {207205} (\bibinfo {year} {2003})}\BibitemShut {NoStop}%
\bibitem [{\citenamefont {Bruin}\ \emph {et~al.}(2013)\citenamefont {Bruin},
  \citenamefont {Borzi}, \citenamefont {Grigera}, \citenamefont {Rost},
  \citenamefont {Perry},\ and\ \citenamefont {Mackenzie}}]{Bruin2013}%
  \BibitemOpen
  \bibfield  {author} {\bibinfo {author} {\bibfnamefont {J.~A.~N.}\
  \bibnamefont {Bruin}}, \bibinfo {author} {\bibfnamefont {R.~A.}\ \bibnamefont
  {Borzi}}, \bibinfo {author} {\bibfnamefont {S.~A.}\ \bibnamefont {Grigera}},
  \bibinfo {author} {\bibfnamefont {A.~W.}\ \bibnamefont {Rost}}, \bibinfo
  {author} {\bibfnamefont {R.~S.}\ \bibnamefont {Perry}}, \ and\ \bibinfo
  {author} {\bibfnamefont {A.~P.}\ \bibnamefont {Mackenzie}},\ }\href {\doibase
  10.1103/PhysRevB.87.161106} {\bibfield  {journal} {\bibinfo  {journal} {Phys.
  Rev. B}\ }\textbf {\bibinfo {volume} {87}},\ \bibinfo {pages} {161106(R)}
  (\bibinfo {year} {2013})}\BibitemShut {NoStop}%
\bibitem [{\citenamefont {Sazonov}\ \emph {et~al.}(2010)\citenamefont
  {Sazonov}, \citenamefont {Gukasov}, \citenamefont {Mirebeau}, \citenamefont
  {Cao}, \citenamefont {Bonville}, \citenamefont {Grenier},\ and\ \citenamefont
  {Dhalenne}}]{Sazonov2010field}%
  \BibitemOpen
  \bibfield  {author} {\bibinfo {author} {\bibfnamefont {A.~P.}\ \bibnamefont
  {Sazonov}}, \bibinfo {author} {\bibfnamefont {A.}~\bibnamefont {Gukasov}},
  \bibinfo {author} {\bibfnamefont {I.}~\bibnamefont {Mirebeau}}, \bibinfo
  {author} {\bibfnamefont {H.}~\bibnamefont {Cao}}, \bibinfo {author}
  {\bibfnamefont {P.}~\bibnamefont {Bonville}}, \bibinfo {author}
  {\bibfnamefont {B.}~\bibnamefont {Grenier}}, \ and\ \bibinfo {author}
  {\bibfnamefont {G.}~\bibnamefont {Dhalenne}},\ }\href {\doibase
  10.1103/PhysRevB.82.174406} {\bibfield  {journal} {\bibinfo  {journal} {Phys.
  Rev. B}\ }\textbf {\bibinfo {volume} {82}},\ \bibinfo {pages} {174406}
  (\bibinfo {year} {2010})}\BibitemShut {NoStop}%
\bibitem [{\citenamefont {Kenzelmann}(2019)}]{Kenzelmann2019}%
  \BibitemOpen
  \bibfield  {author} {\bibinfo {author} {\bibfnamefont {M.}~\bibnamefont
  {Kenzelmann}},\ }\href@noop {} {\bibfield  {journal} {\bibinfo  {journal}
  {private communication}\ } (\bibinfo {year} {2019})}\BibitemShut {NoStop}%
\bibitem [{\citenamefont {Katsufuji}\ and\ \citenamefont
  {Takagi}(2004)}]{Katsufuji2004magnetocapacitance}%
  \BibitemOpen
  \bibfield  {author} {\bibinfo {author} {\bibfnamefont {T.}~\bibnamefont
  {Katsufuji}}\ and\ \bibinfo {author} {\bibfnamefont {H.}~\bibnamefont
  {Takagi}},\ }\href {\doibase 10.1103/PhysRevB.69.064422} {\bibfield
  {journal} {\bibinfo  {journal} {Phys. Rev. B}\ }\textbf {\bibinfo {volume}
  {69}},\ \bibinfo {pages} {064422} (\bibinfo {year} {2004})}\BibitemShut
  {NoStop}%
\bibitem [{\citenamefont {Saito}\ \emph {et~al.}(2005)\citenamefont {Saito},
  \citenamefont {Higashinaka},\ and\ \citenamefont
  {Maeno}}]{Saito2005magnetodielectric}%
  \BibitemOpen
  \bibfield  {author} {\bibinfo {author} {\bibfnamefont {M.}~\bibnamefont
  {Saito}}, \bibinfo {author} {\bibfnamefont {R.}~\bibnamefont {Higashinaka}},
  \ and\ \bibinfo {author} {\bibfnamefont {Y.}~\bibnamefont {Maeno}},\ }\href
  {\doibase 10.1103/PhysRevB.72.144422} {\bibfield  {journal} {\bibinfo
  {journal} {Phys. Rev. B}\ }\textbf {\bibinfo {volume} {72}},\ \bibinfo
  {pages} {144422} (\bibinfo {year} {2005})}\BibitemShut {NoStop}%
\bibitem [{\citenamefont {Khomskii}(2012)}]{Khomskii2012electric}%
  \BibitemOpen
  \bibfield  {author} {\bibinfo {author} {\bibfnamefont {D.~I.}\ \bibnamefont
  {Khomskii}},\ }\href {https://www.nature.com/articles/ncomms1904} {\bibfield
  {journal} {\bibinfo  {journal} {Nat. Commun.}\ }\textbf {\bibinfo {volume}
  {3}},\ \bibinfo {pages} {904} (\bibinfo {year} {2012})}\BibitemShut {NoStop}%
\end{thebibliography}

%

\end{document}